\documentclass[aps,prd,superscriptaddress,amsfonts,amssymb,amsmath,showpacs,twocolumn,nofootinbib,10pt]{revtex4-2}
\usepackage{bm}
\usepackage{amsfonts}
\usepackage{latexsym}
\usepackage{graphicx}
\usepackage{palatino}
\usepackage{xcolor} 
\usepackage{mathpazo}
\usepackage{tensor}
\usepackage{textcomp}
\usepackage{booktabs}
\usepackage{dcolumn}
\usepackage{booktabs}
\usepackage{multirow}
\usepackage{hyperref}
\hypersetup{colorlinks,citecolor=blue}
\usepackage{amsmath}
\usepackage{xcolor}
\usepackage{orcidlink}
\usepackage{epsfig}
\usepackage{caption}
\usepackage{subcaption}
\usepackage{commath}
\hypersetup{colorlinks,citecolor=blue}
\hypersetup{colorlinks=true,linkcolor=magenta,filecolor=magenta,    urlcolor=blue}

\def\be{\begin{equation}}
\def\ee{\end{equation}}
\def\bea{\begin{Eqarray}}
\def\eea{\end{Eqarray}}

\usepackage{enumitem}

\begin{document}

\title{\bf Compact Objects in 4D Einstein Gauss Bonnet Gravity: A Data Based Perspective}

\author{Puja Mukherjee~\orcidlink{0009-0006-4997-1600}}
\email{pmukherjee967@gmail.com} 
\affiliation{Department of Mathematics, Indian Institute of Engineering Science and Technology, Shibpur, Howrah-711 103, India}
\author{Himanshu Chaudhary~\orcidlink{0000-0002-6376-0707}}
\email{himanshu.chaudhary@ubbcluj.ro,\\
himanshuch1729@gmail.com}
\affiliation{Department of Physics, Babeș-Bolyai University, Kogălniceanu Street, Cluj-Napoca, 400084, Romania}
\affiliation{Research Center of Astrophysics and Cosmology, Khazar University, Baku, 41 Mehseti Street, AZ1096, Azerbaijan}
\author{Ujjal Debnath~\orcidlink{0000-0002-2124-8908}}
\email{ujjaldebnath@gmail.com} 
\affiliation{Department of Mathematics, Indian Institute of Engineering Science and Technology, Shibpur, Howrah-711 103, India}
\author{G.Mustafa~\orcidlink{0000-0003-1409-2009}}
\email{gmustafa3828@gmail.com}
\affiliation{Department of Physics,
Zhejiang Normal University, Jinhua 321004, People’s Republic of China}
\begin{abstract}
Cosmic evolution is the most sensational topic among researchers of modern cosmology. This article explores cosmic evolution in 4D Einstein Gauss Bonnet gravity, focusing on mass accretion of compact objects (black holes and wormholes) by dark energy. Three DE models CPL, JBP, and BA parameterizations are studied within 4D EGB gravity, with their Hubble parameters derived and compared against observational data (Cosmic Chronometers, Type Ia Supernovae, and Baryon Acoustic Oscillations). Bayesian analysis favors the CPL and BA models, with CPL providing the best fit. For black holes, mass accretion of CPL and JBP DE shows transitions between quintessence and phantom eras, while BA and $\Lambda$CDM strictly exhibit quintessence-like behavior, driving cosmic acceleration. In contrast, wormholes exhibit the opposite trend, favoring a phantom-dominated era for the BA and $\Lambda$CDM models. The study highlights the dynamic nature of DE in 4D EGB gravity and its role in cosmic expansion.
\end{abstract}

\maketitle

\section{Introduction:}\label{In}
Gravity, a governing force that is nothing more than the curvature of spacetime caused by mass and energy, is one of the most powerful forces known to exist in this universe. It controls almost everything from structure formation to cosmic evolution, as well as the expansion of the universe. Now, just like when there is darkness, light will also be somewhere..... in our universe, there also exist some anti-gravity elements that counter the effects of gravity, leading to the current accelerated expansion phase of the universe. The prime suspect for this is an energy that looms over the universe like a dark cloud and exhibits strong repulsion to gravity, paving the way for space to expand faster with the formal name as ``Dark energy". Its association with vacuum energy leads to a dark energy model with a constant equation of state parameter known to be the famous cosmological constant~($\Lambda$) \cite{peebles2003cosmological}. A benchmark was set in the field of modern cosmology by combining the cosmological constant with the cold dark matter~(CDM), unveiling the concept of the standard~$\Lambda$CDM model of the universe. This is so far the most compatible candidate of the dark-energy models, which can easily accommodate problems regarding fine-tuning or cosmic coincidence \cite{weinberg1989cosmological,padmanabhan2003cosmological}. But in spite of its theoretical brilliance, it faces problems when it comes to the observational relevance, like the ``Hubble tension" or the ``$S_8$ tension". So, the need for more compliant theories arises from here. One of such theories is to replace the usual equation of state parameter of dark energy with a function of redshift variable~$z$, a process commonly known as the dark energy parameterization method or redshift parameterization method~i.e., $p(z)=\omega(z)\rho(z)$, which is quite consistent with the principles of usual general relativity. This parameterization technique can very well incorporate accurate measurements of the equation of state parameter and how it will evolve through cosmic evolution, revealing some of the innate and dynamical properties of the mysterious dark energy \cite{copeland2006dynamics,frieman2008dark}. Several such dark-energy parameterizations are available in the literature, among which we consider three different types of widely used parameterization: Chevallier-Polarski-Linder
\cite{chevallier2001accelerating,linder2003exploring,linden2008test}, Jassal-Bagla-Padmanabhan \cite{bagla2003cosmology,jassal2005wmap}~and~Barboza-Alcaniz \cite{barboza2008parametric} in this particular work.

Another theory is built on the modification of general relativity, creating different versions of modified gravity \cite{tsujikawa2010modified}, which are quite favorable to the current accelerated expansion phase of the universe and to the modern cosmological data. This concept can play a vital role in the field of quantum gravity theory due to its ability to influence the whole gravitational phenomenon. We can very well explore the idea of gravitational interaction through higher-dimensional gravity theories. An interesting extension of the 4-dimensional Einstein gravity theory is the well-known Einstein–Gauss–Bonnet gravity theory in dimension~($D> 4$),~while for $D=4$~it is supposed to be trivial, as by the famous Lovelock theory, the Gauss-Bonnet term vanishes in this particular case \cite{lovelock1971einstein}. But in 2020, this concept changed as a correction to this theory by introducing a rescaling term of the Gauss–Bonnet coupling parameter~($\alpha$) through~$\alpha\to\frac{\alpha}{D-4}$~for the limit~$D-4\to0$~was introduced \cite{glavan2020einstein}. This theoretical approach, famously known as the 4-dimensional Einstein–Gauss–Bonnet~(EGB)~gravity theory. The spherically symmetric black hole solutions in this 4D EGB gravity were also obtained in \cite{glavan2020einstein}. Later, several other generalized versions of black hole solutions were created and observed for this particular gravity theory \cite{ghosh2020generating,jusufi2020nonlinear,zhang2021bardeen,fernandes20224d,fernandes2020charged,rayimbaev2022magnetized,rayimbaev2022shadow,kumar2022bardeen}.

Compact objects like black holes and wormholes are the center of attraction for many cosmologists as they can exhibit some fascinating features, and provide us with a chance to validate the observational data. Although black holes are quite simple in nature and come together with an event horizon, wormholes can be quite complex in their geometry and topological structure \cite{morris1988wormholes,morris1988wormhole,visser1995lorentzian}. Basically, wormholes are imaginary corridors that connect distant points in space-time or two different universes. Numerous works on different wormhole solutions and their properties in 4D EGB gravity were explored in \cite{jusufi2020wormholes,godani2022stability,hassan2024possibility}. Previously, many research studies were conducted on the concept of the mass accretion process of compact objects \cite{john2013accretion,debnath2015accretions,debnath2014accretions,debnath2020nature,bandyopadhyay2021accretions,mukherjee2023accretion}, as mass is an essential property of these objects that can reveal some noteworthy details about their evolution. In this study, we focus our research on the mass accretion phenomena of such compact objects to investigate their evolution in the framework of the 4D EGB gravity theory. Mainly, modified gravities are made by changing the gravitational laws in such a way that they can very well explain the accelerating expansion of our universe without considering any extra dark energy candidate. But in this era of modern cosmology, it is an effective approach to combine modified gravity with dark energy to get more enthralling results. We follow the same path of incorporating 4D EGB gravity and dark-energy parameterizations in this work.\\\\ 
In addition, to achieve results that are compatible with observation data, many researchers use the technique of constraining the free parameters. Several such datasets: The CMB data set \cite{ade2014planck,adam2016planck}, SNLS3 \cite{guy2010supernova}, combined data from BAO, HST, and Union 2.1 \cite{li2013planck}, the recently introduced Data Release 1 of the DESI BAO measurements \cite{adame2025desi,adame2025desis,adame2025desii}, etc. are available for this purpose. In fact, parameter constraining of different dark energy parameterization models \cite{paul2013observational,barboza2008parametric,biswas2015observational,huang2009fitting,feng2012new,paul2012modified,liu2008revisiting} and application of this technique in the mass accretion process in different modified gravity theories \cite{basak2025accretion,mukherjee2024constraining,mukherjee2025parameter} are quite popular nowadays. This works as a motivating factor for us to design our current study based on this parameter-constraining technique.

The outline of this work is presented as: in Sec.\ref{EGB_thory}, we discuss the concept, basic equations, together with the modified Friedmann equations as well as the black hole and wormhole solutions on the 4D EGB gravity theory. Next, in Sec.\ref{DE_parametrization} we consider the proposed dark-energy parameterizations and their corresponding density equation, and also derive the associated Hubble parameter equations sequentially. We constrain the free parameters by the Markov Chain Monte Carlo technique and compare the proposed models from the perspective of data in Sec.\ref{Data}. Then we derive the mass functions in terms of redshift and depict them through graphs with an elaborated explanation for both the black hole and wormhole cases, one by one, for all of the proposed models in Sec.\ref{mass_accretion}. Finally, we conclude this work by explaining the acquired results in a detailed manner in Sec.\ref{conclusion}.

\section{4-dimensional Einstein-Gauss-Bonnet (EGB) gravity theory:}\label{EGB_thory}
In this section, we discuss our motivation behind studying 4D EGB gravity, its basic concept, and the associated modified Friedmann equations, together with the black hole and wormhole solutions on this gravity in detail.

\subsection{Preface:}\label{preface}
Einstein-Gauss-Bonnet (EGB) gravity is a theory carefully crafted by considering string theory and quantum gravity together, which is basically a natural extension of usual general relativity. But in 2020 \cite{PhysRevLett.124.081301} the proposal of the 4D EGB gravity theory marked a significant milestone as it introduced the non-trivial Gauss-Bonnet effects in four-dimensional spacetime, where traditionally the Gauss-Bonnet term is considered as a topological invariant, hence not contributing to the dynamics. This concept sparks interest, as it provides a new 4-dimensional modified gravity theory, together with some additional curvature corrections beyond the usual general relativity. Also, unlike some of the recently developed modified gravity theories like $f(R)$ gravity, this theory has a special type of higher derivative terms that omit the possibilities of ghosts \cite{zubair2023bouncing}. For small values of the Gauss-Bonnet coupling parameter, it resembles general relativity, but with stringy or quantum corrections. Even some studies suggested that this gravity can alter late-time cosmology and resolve the issues of singularities \cite{ghosh2020generating,fernandes20224d}. So, despite some mathematical consistency, such as limit $D\to4$ which does not rigorously justify in all cases, leading to strong coupling issues or breaking of diffeomorphism invariance \cite{aoki2020consistent} or the need to introduce new alternative formulations such as Kaluza-Klein reduction \cite{fernandes20224d,ghosh2020generating,tsujikawa2022instability} or scalar-tensor equivalence \cite{banerjee2021quark,cardenas2023scalar} to justify this theory, it is still an active research topic to explore among researchers.
Now, we discuss the mathematical structure of this theory.

\subsection{Basic Concept and Modified Friedmann Equations:}\label{EGB_Equations}
Let us consider the action for the 4-dimensional Einstein-Gauss-Bonnet (EGB) gravity with Gauss-Bonnet coupling constant in its re-scaled form, together with minimally coupled nonlinear electrodynamics given by \cite{kumar2020hayward,rayimbaev2022magnetized,zhang2021bardeen,rayimbaev2022shadow}:
\begin{equation}\label{action.equ}
\begin{split}
S = & \frac{1}{16\pi} \int d^4x \sqrt{-\bar{g}} \left[ R + \alpha \left( R^2 - 4 R_{\mu\nu} R^{\mu\nu} + R_{\mu\nu\zeta\eta} R^{\mu\nu\zeta\eta} \right) \right. \\
   & \left. - 4 \mathcal{L}(F) \right],
\end{split}
\end{equation}
where~$\alpha$~stands for the Gauss-Bonnet coupling parameter and $\mathcal{L}(F)$~is the Lagrangian density, a function of the electromagnetic field invariant $F$, given by $F\equiv F_{\mu\nu}F^{\mu\nu}$~and~$F_{\mu\nu}\equiv\partial_\mu A_\nu-\partial_\nu A_\mu$~is the electromagnetic field tensor term for the gauge potential $A_\mu$. Again, $\bar{g}$~is the value of the determinant of the metric tensor $g_{\mu\nu}$~and~$R$~,~$R_{\mu\nu}$~and~$R^{\mu}_{\nu\zeta\eta}$~are respectively the Ricci scalar, Ricci and Riemann tensors.

The specific form of this Lagrangian density~$\mathcal{L}(F)$~for nonlinear electrodynamics field in the 4-dimensional spacetime is given as \cite{ayon1998regular,ayon2005four,kumar2020hayward,rayimbaev2022shadow,rayimbaev2022magnetized}:
\begin{equation}\label{Lagrangian_densit.equ}
\mathcal{L}(F)=\frac{3\mu}{2g}\left[\frac{\sqrt{2g^2F}}{1+\sqrt{2g^2F}}\right]^\frac{5}{2},   
\end{equation}
together with
\begin{equation}\label{electromagnetic_field.equ}
F=\frac{g^2}{2r^4}.    
\end{equation}
Where~$g$~stands for the magnetic charge governed by the field of nonlinear electrodynamics.\\\\
Now, considering the isotropic, homogeneous and spatially flat Friedmann–Robertson–Walker metric of the universe given by
\begin{equation}\label{FRW_metric}
ds^2=-dt^2+a^2(t)\left(dr^2+r^2\left(d\theta^2+sin^2\theta d\phi^2\right)\right),   
\end{equation}
where $a(t)$~stands for the scale factor of the universe, and the scalar curvature has a value of zero according to our assumption that the model is flat. The Hubble parameter is $H=\frac{\dot{a}}{a}$~and~suppose that the universe of 4-dimensional EGB gravity consists of matter of perfect fluid with the stress-energy tensor of the form $T_{\mu\nu}=\{-\rho,p,p,p,....\}$, where $\rho$~and~$p$~are respectively the energy density and pressure components of the matter field. Then the corresponding modified Friedmann equations are given by \cite{fernandes20224d,sheykhi2023growth,haghani2020growth}:
\begin{equation}\label{F.Eq.1}
 H^2+\alpha H^4=\frac{\kappa^2}{3}\rho,  
\end{equation}
\begin{equation}\label{F.Eq.2}
(1+2\alpha H^2)\dot{H}=-\frac{\kappa^2}{2}(\rho+p).    
\end{equation}\\
The associated matter field obeys the following form of the usual conservation equation, given by
\begin{equation}\label{conservation.equ}
\dot{\rho}+3H(\rho+p)=0.    
\end{equation}

\subsection{Black hole solution in 4-dimensional Einstein-Gauss-Bonnet (EGB) gravity theory:}\label{Black_hole}
The solution of the gravitational field after varying the action given in Eq.(\ref{action.equ}), known as the regular Bardeen black hole in 4D EGB gravity together with non-linear electrodynamics, is given as follows \cite{islam2024strong,rayimbaev2022shadow,kumar2022bardeen,zhang2021bardeen}:
\begin{equation}\label{BH.metric}
 ds^2=-F(r)dt^2+\frac{1}{F(r)}dr^2+r^2(d\Theta^2+sin^2\Theta d\psi^2),   
\end{equation}
in correspondence with the following lapse function, given by
\begin{equation}\label{BH.lapse_function}
F(r)=1+\frac{r^2}{2\alpha}\left(1\pm\sqrt{1+\frac{8M\alpha}{(r^2+g^2)^\frac{3}{2}}}\right),    
\end{equation}
where~$ M$~is the total black hole mass~and~$ g$~is the magnetic charge due to the non-linear electrodynamic field. 
The above metric represents two different branches of the black hole solution for two different signs. For the negative branch, if we consider the far-region condition and take the limit $\alpha\to 0$~,~we have the solution in the form of a regular Bardeen black hole in general relativity with positive gravitational mass \cite{bardeen1968non,rayimbaev2022shadow,rayimbaev2022magnetized,zhang2021bardeen} and also it reduces to the solution of the Gauss-Bonnet-Schwarzschild black hole in general relativity for $g=\alpha=0$ \cite{zhang2021bardeen}. However, the positive branch is unstable due to the negative value of the gravitational mass \cite{boulware1985string}. So, it is feasible for us to consider only the negative branch of the black hole metric for the rest of our calculation. In addition, we have taken the liberty to consider $\frac{\alpha}{M^2}\to \alpha$~\cite{rayimbaev2022shadow} for further investigation on this article.\\\\
So, our required form of lapse function for the Bardeen black hole metric given in Eq.(\ref{BH.metric}) in 4D EGB gravity becomes:
\begin{equation}\label{BH.lapse_function_1}
F(r)=1+\frac{r^2}{2\alpha}\left(1-\sqrt{1+\frac{8M\alpha}{(r^2+g^2)^\frac{3}{2}}}\right).    
\end{equation}

\subsection{Wormhole solution in 4-dimensional Einstein-Gauss-Bonnet (EGB) gravity theory:}\label{Worm_hole}
The wormhole solution in 4D EGB gravity is given by:
\begin{equation}\label{Wh.metric}
ds^2=-e^{2\phi(r)}dt^2+\frac{dr^2}{1-\frac{b(r)}{r}}+r^2(d\Theta^2+sin^2\Theta d\psi^2),
\end{equation}
where ~$\phi(r)$~is the redshift function, which should have a finite value to eliminate the possibility of forming an event horizon. Again,~$b (r)$~is the spatial shape function for the wormhole geometry that gives the idea of the shape of the wormhole in the embedding diagram \cite{mehdizadeh2015einstein}~and~it should satisfy the required boundary condition $b(r=r_0)=r_0$~at the wormhole throat radius for~$r_0\leq r<\infty$. Also, by the traversability condition of the wormhole, $b(r)$ must obey the flaring-out condition that we can easily acquire from the embedding calculation as follows:
\begin{equation}\label{traversability.equ}
\frac{b(r)-r~b'(r)}{b^2(r)}>0~,   
\end{equation}
which in compact form can be written as $b'(r)<1$~at the wormhole throat~$r=r_0$~together with the condition~$1-\frac{b(r)}{r}\geq 0$.\\
Again, let us take a constant redshift function, which means the wormhole solution with zero tidal force~,~thus $\phi(r)=$ constant~,~which gives an exact wormhole solution. Also, the wormhole shape function is given in its explicit form as below \cite{jusufi2020wormholes}:
\begin{equation}\label{Wh.shapefunction}
b(r)=-\frac{r^3(3\omega+1)}{\alpha(3\omega-1)}\left(1\pm\sqrt{1+\frac{4\alpha\mathcal{A}}{r^3r_0}}\right),    
\end{equation}
with 
\begin{equation}\label{Wh.constant}
\mathcal{A}=\frac{(3\omega-1)}{(1+3\omega)^2}\left[r^2_0(3\omega+1)+\alpha(3\omega-1)\right].  
\end{equation}
So, it is clear that Eq.(\ref{Wh.shapefunction}) exists only for $\omega\neq 1/3$, and its~``$\pm$"~sign stands for two different branches of the wormhole solution. As discussed above, EGB black holes with negative branches are stable and free of ghosts, whereas the positive branch is unstable, and their gravitational degree of freedom is a ghost \cite{boulware1985string}. Similarly, for the wormhole solution, whenever $\alpha\to 0$~,~the positive branch gives 
\begin{equation*}
 \frac{b(r)}{r}=-\frac{r^2(3\omega+1)}{(3\omega-1)\alpha}-\frac{r_0}{r}....~,   
\end{equation*}
representing wormhole solutions in an anti-de Sitter/de Sitter universe solely depending on the sign of the Gauss-Bonnet coupling parameter $\alpha$. 
Besides, for the limit $\alpha\to 0$~,~the negative branch gives
\begin{equation*}
 \frac{b(r)}{r}=\frac{r_0}{r}+....~,    
\end{equation*}
Hence, if the cosmological constant disappears, we get the standard Morris–Thorne wormhole from this negative branch \cite{jusufi2020wormholes}. Thus, it is convenient to consider the negative branch of the wormhole solution for further study in this article.
So, our required wormhole metric in 4D EGB gravity becomes:
\begin{equation}\label{Wh_metric.form}
\begin{split}
ds^2 = & -dt^2 + \frac{dr^2}{1 + \frac{r^2(3\omega + 1)}{\alpha(3\omega - 1)}\left( 1 - \sqrt{1 + \frac{4\alpha \mathcal{A}}{r^3r_0}} \right)} \\
& + r^2 \left( d\Theta^2 + \sin^2\Theta d\psi^2 \right).
\end{split}
\end{equation}
Consequently,  the above form of the wormhole metric is valid for $\omega\geq-1/3$~with~$\omega\neq 1/3$. But for $\omega<-1/3$~,~the above solution inverts its sign.\\\\
Now, using the assumption that our universe is a combination of dark energy and dark matter, together with the fact that dark matter is almost pressureless, we can deduce the expression of the energy density for dark matter from Eq.(\ref{conservation.equ}) with the help of the redshift equation~$a=\frac{1}{1+z}$~as given below:
\begin{equation*}\label{energy_density_form.DM}
 \rho_{_{DM}}=\rho_{_{DM0}}(1+z)^3,  
\end{equation*}
where~ $\rho_{_{DM0}}$~stands for the present energy density value of dark matter.
So, in terms of dimensionless density parameter related to dark matter~$\Omega_{_{m0}}=\frac{8\pi G\rho_{_{DM0}}}{3H^2_0}$~,~we get the energy density equation of dark matter in the following form:
\begin{equation}\label{energy_density_equation.DM}
\rho_{_{DM}}=\frac{3H^2_0}{8\pi G}\Omega_{_{m0}}(1+z)^3.    
\end{equation}
Finally, the expression of the corresponding Hubble parameter associated with the 4D EGB gravity is obtained in the following form:
\begin{equation}\label{Hubble_parameter.equation}
\begin{split}
H^2(z) = & \frac{1}{2\alpha} \left[ -1 + \left( 1 + 4\alpha H_0^2 \left\{ \Omega_{_{m0}} (1+z)^3 + \right. \right. \right. \\
& \left. \left. \left. \left( 1 + \alpha H_0^2 - \Omega_{_{m0}} \right) \left( \frac{\rho_{_{DE}}}{\rho_{_{DE0}}} \right) \right\} \right)^{\frac{1}{2}} \right],
\end{split}
\end{equation}
where ~$\Omega_{_{DE0}}=\frac{8\pi G\rho_{_{DE0}}}{3H^2_0}$~~is the dimensionless density parameter related to dark energy,~which satisfies the relation:
\begin{equation}\label{DE_relation}
\Omega_{_{DE0}}+\Omega_{_{m0}}-\alpha H^2_0=1~,   
\end{equation}
and~$\rho_{_{DE0}}$~is the present energy density value of the dark energy.

\section{Dark energy parameterization:}\label{DE_parametrization}
Dark energy parameterization is a generalized approach to explore dark energy dynamically rather than simply assuming a constant value of the equation of state parameter~$\omega$. In the standard $\Lambda$CDM model of the universe (for $\omega=-1$), a constant value of the dark energy density is assumed. In addition, by solving the scalar field equation associated with some particular theoretical model, we can extract some dynamical values of~$\omega$. However, this approach has several drawbacks, such as not ensuring any model-independent way to explore the broader parameter space and not facilitating comparisons among different cosmological models. Different dark energy parameterizations of the equation of state parameter~$\omega(z)$ depending on the redshift function $z$ are proposed to resolve this issue and establish a more potent and generalized perspective. This gives us an effective gateway to capture all the possible potential in variations of $\omega(z)$ over time. Dark energy parameterizations are vital in understanding the cosmic background in which black holes form and grow, as well as their thermodynamics and mass accretion physics. However, it plays an intriguing yet subtle role in the studies of wormholes, specifically in their stability, cosmological viability, and energy conditions. Also, this process significantly reduces the possibility of assumptions leading to data-driven conclusions rather than theoretical prejudices. It is ideal for exploring modified gravity theories and differentiating between competing models with the help of Supernovae (SNIa)~,~Baryon Acoustic Oscillations (BAO)~,~Cosmic Microwave Background (CMB)~,~Weak Lensing observations, and 
to optimize future observational surveys such as~Euclid~,~JWST~,~DESI~,~LSST~etc.\\\\
Mainly, there are two mainstream families of dark-energy parameterization models based on the redshift function, given as follows:
\begin{itemize}
    \item {\bf{Equation of state parameter for family 1~:}}
    \begin{equation}\label{family_1}
   \omega(z)=\omega_0+\omega_1\left(\frac{z}{1+z}\right)^n,     
    \end{equation}
    \end{itemize}
and\\
\begin{itemize}
    \item {\bf{Equation of state parameter for family 2~:}}
    \begin{equation}\label{family_2}
   \omega(z)=\omega_0+\omega_1\frac{z}{(1+z)^n}~~,     
    \end{equation}
    \end{itemize}
where both $\omega_0$~and~$\omega_1$~are two unknown parameters, $n$~denotes a natural number. A conventional and easy approach is to consider~$\omega(z)$~as a linear, first-order function of the redshift variable, taking into account the time variation in the equation of state parameter. This gives the expression as $\omega(z)=\omega_0+\omega_1 z$, where~$\omega_0$~and~$\omega_1$~respectively stand for the present value of $\omega$~and~its variation with redshift. But this particular parameterization has some significant complications, such as, for large values of $z$~, its precision diminishes, which in turn creates a difficulty in deriving precise constraints for the equation of state parameter in the case of specifically CMB distance measurement. So, to resolve these challenges, alternative forms of the equation of the state parameter and its associated distance-redshift relationship are established, which can help to reconstruct the scalar field equations more accurately. On the basis of this fact, several new dark-energy parameterizations are being proposed. In this particular work, we are going to study three such dark-energy parameterization models in a consecutive manner.   

\subsection{Chevallier-Polarski-Linder CPL) Parameterization:}\label{CPL}
 One of the most suitable forms of such parameterizations is obtained when we take $n=1$ for both the families $1$~and~$2$~(see Eq.(\ref{family_1})~and~Eq.(\ref{family_2}))~leading to the same parameterization form in the terms of the redshift function given as follows \cite{biswas2015observational,mehrabi2018growth,kumar2025exploring}:
\begin{equation}\label{EOS_CPL}
\omega_{_{CPL}}(z)=\omega_0+\omega_1\left(\frac{z}{1+z}\right),   
\end{equation}
where the parameter~$\omega_1$~controls the evolution over time of the equation of state parameter.
This ansatz was first described in \cite{chevallier2001accelerating}~and then studied in a more elaborate way in \cite{linder2003exploring}. This parametrization technique is named after Chevallier-Polarski-Linder as ``CPL parameterization". It is basically the Taylor series expansion
of the dark energy equation of state parameter~$\omega(z)$~with respect to the redshift function $z$ up to the first order. At the early universe~($z\to\infty$)~and~present epochs~($z=0$)~CPL behave quite well but diverge at the future time~($z=-1$). Many significant literature supporting the CPL parametrization as a nice candidate to preserve the dynamics of different dark energy models, specifically step-like ones, is available to look at \cite{linden2008test}.\\
The energy density equation for the CPL parameterization deduced from the dark energy conservation equation is given as \cite{malekjani2025cosmological,mehrabi2018growth,biswas2015observational}:
\begin{equation}\label{energy_density_CPL}
\rho_{_{_{CPL}}}(z)=\rho_{_{_{CPL0}}}\left(1+z\right)^{3\left(1+\omega_0+\omega_1\right)}exp\left(\frac{-3\omega_1 z}{1+z}\right),    
\end{equation}
where $\rho_{_{CPL0}}$~gives the present energy density value for the CPL parametrization.\\
Now, from Eq.(\ref{Hubble_parameter.equation}) we can formulate the desired expression of the Hubble parameter for the CPL parameterization technique in the context of 4D EGB gravity as below:
\begin{equation}\label{Hubble_parameter_CPL}
\begin{split}
H^2_{_{CPL}}(z) = & \frac{1}{2\alpha} \left[ -1 + \left( 1 + 4\alpha H_0^2 \left\{ \Omega_{_{m0}} (1+z)^3 + \right. \right. \right. \\
 & \left. \left. \left. \left( 1 + \alpha H_0^2 - \Omega_{_{m0}} \right) \left( 1 + z \right)^{3(1 + \omega_0 + \omega_1)} \right. \right. \right. \\
\times & \left. \left. \left. \exp\left( \frac{-3\omega_1 z}{1+z} \right) \right\} \right)^{\frac{1}{2}} \right].
\end{split}
\end{equation}
\subsection{Jassal-Bagla-Padmanabhan (JBP) Parameterization:}\label{JBP}
This parameterization technique belongs to family 2,~as we take $n=2$~ in Eq.(\ref{family_2})~we can formulate the expression of the equation of state parameter as a function of the redshift given by \cite{jassal2005wmap,kumar2025exploring,biswas2015observational,bandyopadhyay2019parameterizing,pantazis2016comparison,rebouccas2024investigating,jassal2005observational}: 
\begin{equation}\label{EOS_JBP}
\omega_{_{JBP}}(z)=\omega_0+\omega_1\frac{z}{(1+z)^2},    
\end{equation}
where both $\omega_0$~and~$\omega_1$~are constants to be chosen from observational analysis. Also, we can observe that this parameterization consists of both linear and quadratic terms of the redshift function. It shows some differences from the CPL parameterization model at low redshifts, as in the current epoch, its term $-\frac{\omega_1}{(1+z)^2}$ becomes comparable to the term $\frac{\omega_1}{(1+z)}$. It is called Jassal-Bagla-Padmanabhan parameterization or ``JBP parameterization" \cite{jassal2005wmap}. Basically, this parameterization technique can avoid the strong divergence in $z\to -1$, that is, in future evolution, making it a better candidate than the CPL parameterization technique. In addition, it can provide a better symmetric evolution around $z \sim 0$. Note that both the CPL and JBP parameterizations reduce to one for the particular value of $\omega_1=0$~,~which can be treated as a special case. 
The energy density equation for the JBP parameterization, deduced from the dark energy conservation equation, is given as \cite{biswas2015observational,chaudhary2024early}:
\begin{equation}\label{energy_density_JBP}
\rho_{_{_{JBP}}}(z)=\rho_{_{_{JBP0}}}\left(1+z\right)^{3\left(1+\omega_0\right)}exp\left(\frac{3\omega_1 z^2}{2(1+z)^2}\right),    
\end{equation}
where $\rho_{_{_{JBP0}}}$~gives the present energy density value for the JBP parametrization.\\
Thus, from Eq.(\ref{Hubble_parameter.equation}) we can formulate the required expression of the Hubble parameter for the JBP parameterization technique in the context of 4D EGB gravity as below:
\begin{equation}\label{Hubble_parameter_JBP}
\begin{split}
H^2_{_{JBP}}(z) = & \frac{1}{2\alpha} \left[ -1 + \left( 1 + 4\alpha H_0^2 \left\{ \Omega_{_{m0}} (1+z)^3 + \right. \right. \right. \\
 & \left. \left. \left. \left( 1 + \alpha H_0^2 - \Omega_{_{m0}} \right) \left( 1 + z \right)^{3(1 + \omega_0 )} \right. \right. \right. \\
\times & \left. \left. \left. \exp\left( \frac{3\omega_1 z^{2}}{2(1+z)^{2}} \right) \right\} \right)^{\frac{1}{2}} \right].
\end{split}
\end{equation}
\subsection{Barboza-Alcaniz (BA) Parameterization:}\label{BA}
In most cases of dark energy clustering, generally~$\omega$~is considered to be~$\omega=$~constant + cold dark matter together with CPL parametrization, but its main drawback is that it cannot justify the effects of dark energy clustering on dark matter perturbations in different parameterizations mainly if the evolution of the equation of state parameter is different from CPL parametrization. So, we need a generalized version of the CPL parameterization, which can be written as \cite{malekjani2025cosmological}:
\begin{equation}\label{generalized_CPL}
\omega_{_{DE}}(z)=\omega_0+\omega_1\frac{z(1+z)^{n-1}}{1+z^n} .   
\end{equation}
Now, taking $n=1$~we can recover the CPL model, whereas for $n=2$~we get the equation of state parameter in terms of the redshift function for a new parameterization model, given by \cite{barboza2008parametric,chaudhary2024addressing,mehrabi2018growth,malekjani2025cosmological}:
\begin{equation}\label{EOS_BA}
\omega_{_{BA}}(z)=\omega_0+\omega_1\frac{z(1+z)}{1+z^2} ,   
\end{equation}
where~$\omega_0$~and~$\omega_1$~are two free parameters. This model is called the Barboza-Alcaniz parameterization or ``BA parameterization" model. It shows a linear behavior for low redshifts, as well as remains well behaved at all times, despite allowing deviation from the usual CPL method, unlike the CPL parameterization method itself, which shows rapid expansion whenever the redshift goes to infinity, i.e., for $\omega_1>0$. BA parameterization has a different redshift dependency, and thus, it can help us to analyze changes in the effects of dark energy clustering for different parametrization methods.
The energy density equation for the BA parameterization, deduced from the dark energy conservation equation, is given as \cite{chaudhary2024addressing,mehrabi2018growth,malekjani2025cosmological}:
\begin{equation}\label{energy_density_BA}
\rho_{_{_{BA}}}(z)=\rho_{_{_{BA0}}}\left(1+z\right)^{3\left(1+\omega_0\right)}\left(1+z^2\right)^{\frac{3\omega_1}{2}},    
\end{equation}
where $\rho_{_{_{BA0}}}$~gives the present energy density value of the BA parametrization.\\
Finally, from Eq.(\ref{Hubble_parameter.equation}) we can formulate the desired expression of the Hubble parameter for the JBP parameterization technique in the context of 4D EGB gravity as below:
\begin{equation}\label{Hubble_parameter_BA}
\begin{split}
H^2_{_{BA}}(z) = & \frac{1}{2\alpha} \left[ -1 + \left( 1 + 4\alpha H_0^2 \left\{ \Omega_{_{m0}} (1+z)^3 + \right. \right. \right. \\
 & \left. \left. \left. \left( 1 + \alpha H_0^2 - \Omega_{_{m0}} \right) \left( 1 + z \right)^{3(1 + \omega_0 )} \right. \right. \right. \\
\times & \left. \left. \left. \left( 1+z^2 \right)^{\frac{3\omega_1}{2}} \right\} \right)^{\frac{1}{2}} \right].
\end{split}
\end{equation}
\section{Methodology and Datasets}\label{Data}
Nested sampling is a Bayesian inference technique designed to efficiently compute the Bayesian evidence* (marginal likelihood), $\mathcal{Z} = p(D|M)$, and simultaneously sample from the posterior distribution. Given data $D$ and model parameters $\theta$, the Bayesian evidence is defined as
$$
\mathcal{Z} = \int L(D|\theta) \, \pi(\theta) \, d\theta,
$$
where $L(D|\theta)$ is the likelihood function and $\pi(\theta)$ is the prior distribution. Nested sampling transforms this multidimensional integral into a one dimensional integral over the prior volume, allowing for efficient evaluation even in high dimensional parameter spaces. This approach not only provides parameter constraints but also enables rigorous model comparison through the evidence values. In this study, we apply nested sampling via the \textit{PyPolyChord} library \textcolor{red}{\cite{handley2015polychord}}\footnote{\url{[https://github.com/PolyChord/PolyChordLite}}. PyPolyChord generates posterior samples and computes the Bayesian evidence while exploring complex likelihood surfaces. We focus on constraining three popular dark energy parameterizations within the 4D Einstein-Gauss–Bonnet Gravity framework:
\begin{itemize}
    \item CPL parameterization
    \item JBP parameterization
    \item BA parameterization
\end{itemize}
For each model, we impose uniform priors on the following cosmological parameters: $H_0 \in [50, 100]$ km s$^{-1}$ Mpc$^{-1}$, $\Omega_{m0} \in [0, 1]$, $\alpha \in [0, 0.1]$, $\omega_0 \in [-3, 1]$, $r_d \in [100, 200]$ for all DE parametrization models and $\omega_1 \in [-3, 2]$ for CPL and BA and $\omega_1 \in [-3, 4]$ for the JBP model. These priors are implemented through PyPolyChord’s \texttt{UniformPrior} class. We configure PyPolyChord with approximately 100 live points and enable clustering to ensure efficient sampling of multimodal and degenerate posterior distributions. The algorithm produces posterior samples for parameter estimation and Bayesian evidence values for model selection. To analyze and visualize the results, we use the \textit{getdist} package \textcolor{red}{\cite{lewis2025getdist}}\footnote{\url{[https://getdist.readthedocs.io/en/latest/plot\_gallery.html}}, which facilitates detailed marginalized posterior distributions and correlation plots. To determine the values of the model parameters, we use several different observational datasets. These include measurements from Cosmic Chronometers, Type Ia Supernovae, and Baryon Acoustic Oscillations. For each dataset, we carefully build likelihood functions that take into account the uncertainties and correlations in the data. Below, we explain each dataset in more detail and describe how their likelihoods are constructed.

\begin{itemize}
    \item \textbf{Cosmic Chronometers:} We consider 15 Hubble measurements spanning from redshift range of $0.1791 \leq z \leq 1.965$, \cite{moresco2012new,moresco2015raising,moresco20166}, obtained via the differential age method. This model-independent technique uses passively evolving galaxies at nearby redshifts to estimate $H(z)$ from the derivative $\Delta z/\Delta t$ \cite{jimenez2002constraining}. To constrain the parameters of the CPL, JBP, and BA parametrizations in the 4D Einstein Gauss Bonnet framework, we use the CC likelihood provided by Moresco in his GitLab repository,\footnote{\protect\url{https://gitlab.com/mmoresco/CCcovariance}} which includes the full covariance matrix accounting for both statistical and systematic uncertainties \cite{moresco2018setting,moresco2020setting}.
    \item \textbf{Type Ia supernova :}
    We also use the Pantheon$^{+}$ dataset (without SHOES calibration), which comprises 1701 light curves from 1550 Type Ia Supernovae (SNe Ia) spanning a redshift range of \( 0.001 \leq z \leq 2.3 \) \cite{brout2022pantheon}. During our analysis, we have excluded SNe Ia measurements below $z < 0.01$, as such low-redshift data are affected by significant systematic uncertainties arising from peculiar velocities. Our analysis follows the likelihood formalism provided in the following GitHub link\footnote{\url{https://github.com/PantheonPlusSH0ES/DataRelease}}, incorporating the total covariance matrix, \(\mathbf{C}_{\text{total}} = \mathbf{C}_{\text{stat}} + \mathbf{C}_{\text{sys}}\), which accounts for both statistical and systematic uncertainties \cite{conley2010supernova}. The likelihood function is expressed as: $\mathcal{L}_{\text{SNe Ia}} = e^{\left(-\frac{1}{2} \Delta\mu^T \mathbf{C}^{-1}_{\text{total}} \Delta\mu \right)},$ where the residual vector $\Delta\mu$ represents the difference between observed and theoretical distance moduli: $\Delta\mu_i = \mu_{\text{obs}}(z_i) - \mu_{\text{th}}(z_i, \Theta).$ The theoretical distance modulus is given by: $\mu_{\text{th}}(z) = 5 \log_{10} \left( \frac{d_L(z)}{\text{Mpc}} \right) + \mathcal{M} + 25,$ where the luminosity distance in a flat FLRW cosmology is: $d_L(z) = c(1 + z) \int_0^z \frac{dz'}{H(z')}.$ This formulation reflects the degeneracy between the nuisance parameter \(\mathcal{M}\) and the Hubble constant \(H_0\). In our analysis, we marginalize over the absolute magnitude parameter $\mathcal{M}$ ; for more details, see Equations~(A9–A12). in~\cite{goliath2001supernovae}).\\\\
    
    \item \textbf{Baryon Acoustic Oscillation :} We also consider 13 Baryon Acoustic Oscillation (BAO) measurements from the Dark Energy Spectroscopic Instrument (DESI) Data Release 2 (DR2) \cite{karim2025desi}. These measurements are derived from various tracers, including the Bright Galaxy Sample (BGS), Luminous Red Galaxies (LRG1, LRG2, LRG3), Emission Line Galaxies (ELG1 and ELG2), Quasars (QSO), and Lyman-$\alpha$ forests.\footnote{\url{https://github.com/CobayaSampler/bao_data}}. The BAO scale is determined by the sound horizon at the drag epoch ($z_d \approx 1060$), given by $r_d = \int_{z_d}^{\infty} \frac{c_s(z)}{H(z)} \, dz$, where $c_s(z)$ depends on the baryon-to-photon ratio. In flat $\Lambda$CDM, $r_d = 147.09 \pm 0.26$ Mpc \cite{collaboration2020planck}. In our analysis, we treat $r_d$ as a free parameter \cite{pogosian2020recombination,jedamzik2021reducing,pogosian2024consistency,lin2021early,vagnozzi2023seven}. To analyze the BAO measurements, we compute three key distance scales: the Hubble distance $D_H(z) = \frac{c}{H(z)}$, the comoving angular diameter distance $D_M(z) = c \int_0^z \frac{dz'}{H(z')}$, and the volume-averaged distance $D_V(z) = \left[ z D_M^2(z) D_H(z) \right]^{1/3}$. To constrain the model parameters, we analyze relevant ratios such as \( D_M(z)/r_d \), \( D_H(z)/r_d \), and \( D_V(z)/r_d \). The BAO likelihood takes the standard Gaussian form: $\mathcal{L}_{\mathrm{BAO}} \propto \exp\left( -\frac{1}{2} \chi^2_{\mathrm{BAO}} \right),$ where the chi-squared is given by $\chi^2_{\mathrm{BAO}} = \Delta \mathbf{D}^\intercal \, \mathbf{C}^{-1} \, \Delta \mathbf{D},$ with $\Delta \mathbf{D} = \mathbf{D}_{\mathrm{obs}} - \mathbf{D}_{\mathrm{th}}$ representing the difference between observed and theoretical BAO distance measures, and $\mathbf{C}^{-1}$ the inverse covariance matrix \footnote{\url{https://github.com/CobayaSampler/bao_data/blob/master/desi_bao_dr2/desi_gaussian_bao_ALL_GCcomb_cov.txt}}.\\\\
\end{itemize}
The posterior distributions of each case in the 4D Einstein-Gauss–Bonnet framework is obtained by maximizing the total likelihood function, expressed as: $\mathcal{L}_{\text{tot}} = \mathcal{L}_{\text{CC}} \times \mathcal{L}_{\text{SNe Ia}}\times \mathcal{L}_{\text{BAO}}$.

Further, we compute the Bayes factor \cite{trotta2008bayes}, defined as $B_{i0} = \frac{p(d|M_i)}{p(d|M_0)}$, where $p(d|M_i)$ and $p(d|M_0)$ represent the Bayesian evidences for model $M_i$ and the reference model $M_0$, respectively. In this analysis, $M_0$ corresponds to the $\Lambda$CDM model, while $M_i$ refers to the CPL, JBP , and BA dark energy parametrizations, all considered within the framework of 4D Einstein–Gauss–Bonnet gravity. Since analytical calculation of the Bayesian evidence is difficult, we use PolyChord, which computes it numerically through a nested sampling algorithm designed to efficiently explore high dimensional parameter spaces. We report the natural logarithm of the Bayes factor, $\ln \mathcal{Z}_{i,0}$ and interpret the results using Jeffreys' scale \cite{jeffreys1961theory}: $|\ln \mathcal{Z}_{i,0}| < 1$ indicates inconclusive evidence; $1 \leq |\ln \mathcal{Z}_{i,0}| < 2.5$ suggests weak support for model $i$; $2.5 \leq |\ln \mathcal{Z}_{i,0}| < 5$ implies moderate support; and $|\ln \mathcal{Z}_{i,0}| > 5$ represents strong support for model $i$. Negative values of $\ln \mathcal{Z}_{i,0}$ indicate a preference for the reference $\Lambda$CDM model.

\begin{figure*}
\begin{subfigure}{.33\textwidth}
\includegraphics[width=\linewidth]{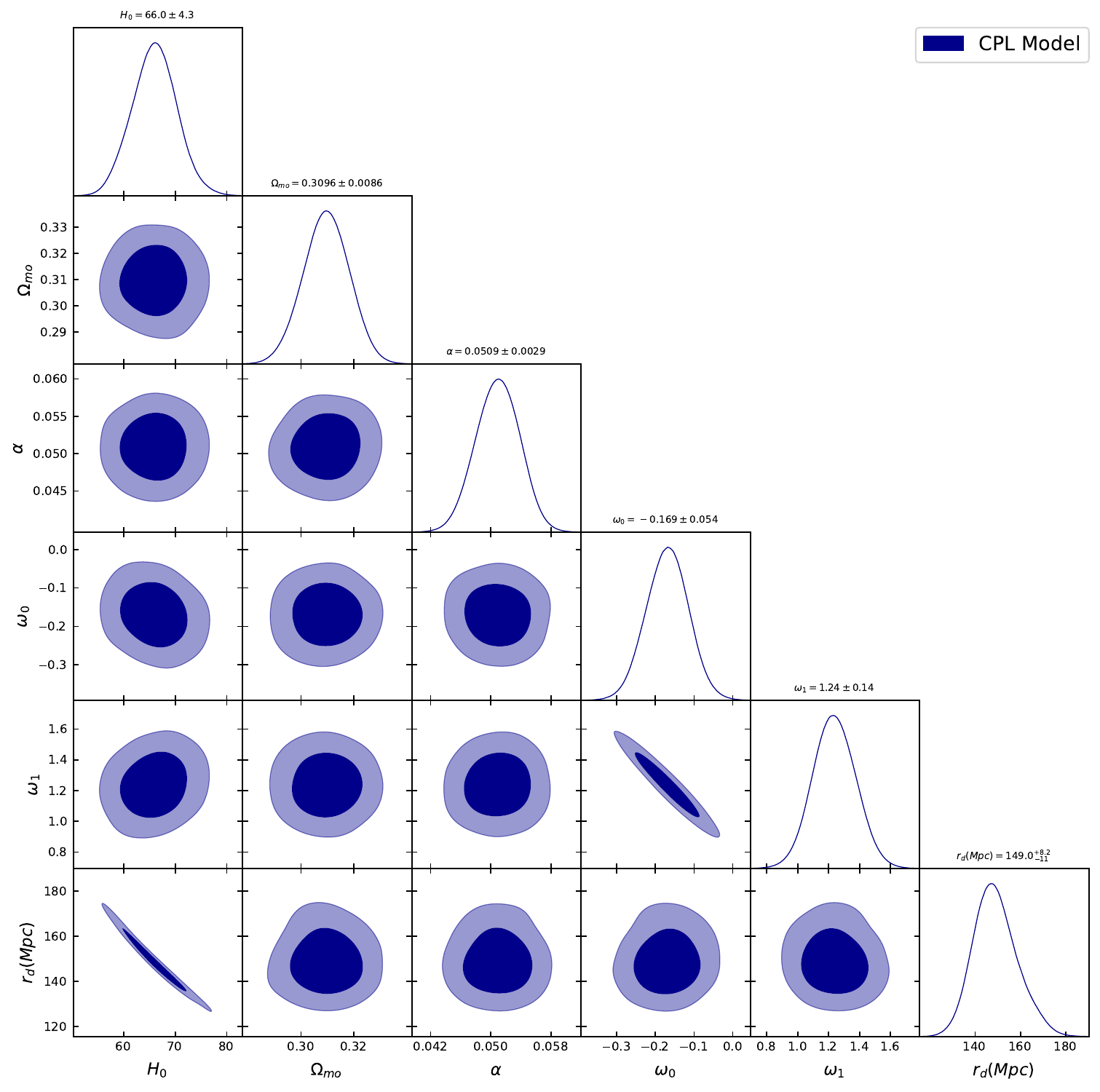}
\end{subfigure}
\hfil
\begin{subfigure}{.33\textwidth}
\includegraphics[width=\linewidth]{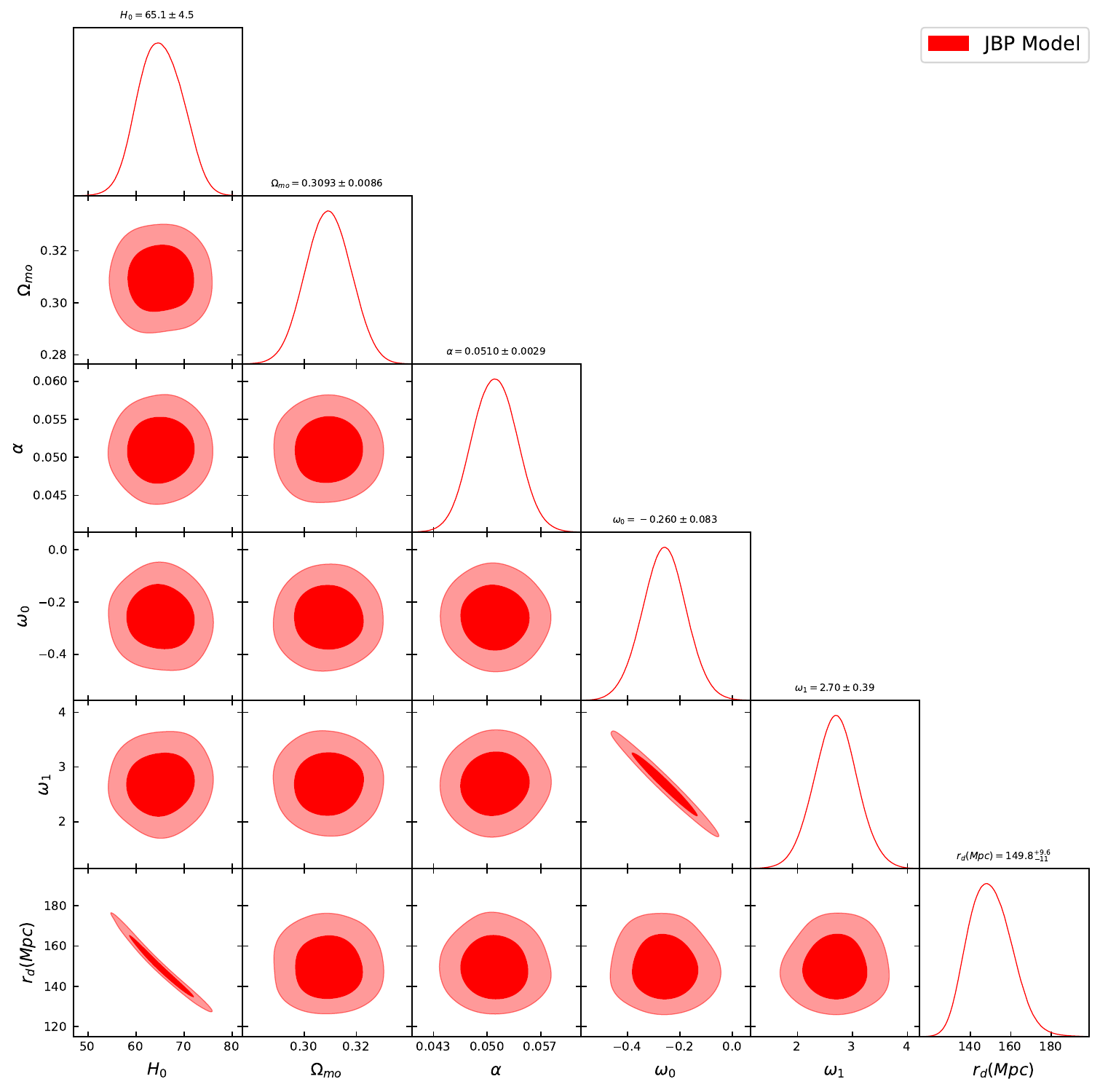}
\end{subfigure}
\hfil
\begin{subfigure}{.33\textwidth}
\includegraphics[width=\linewidth]{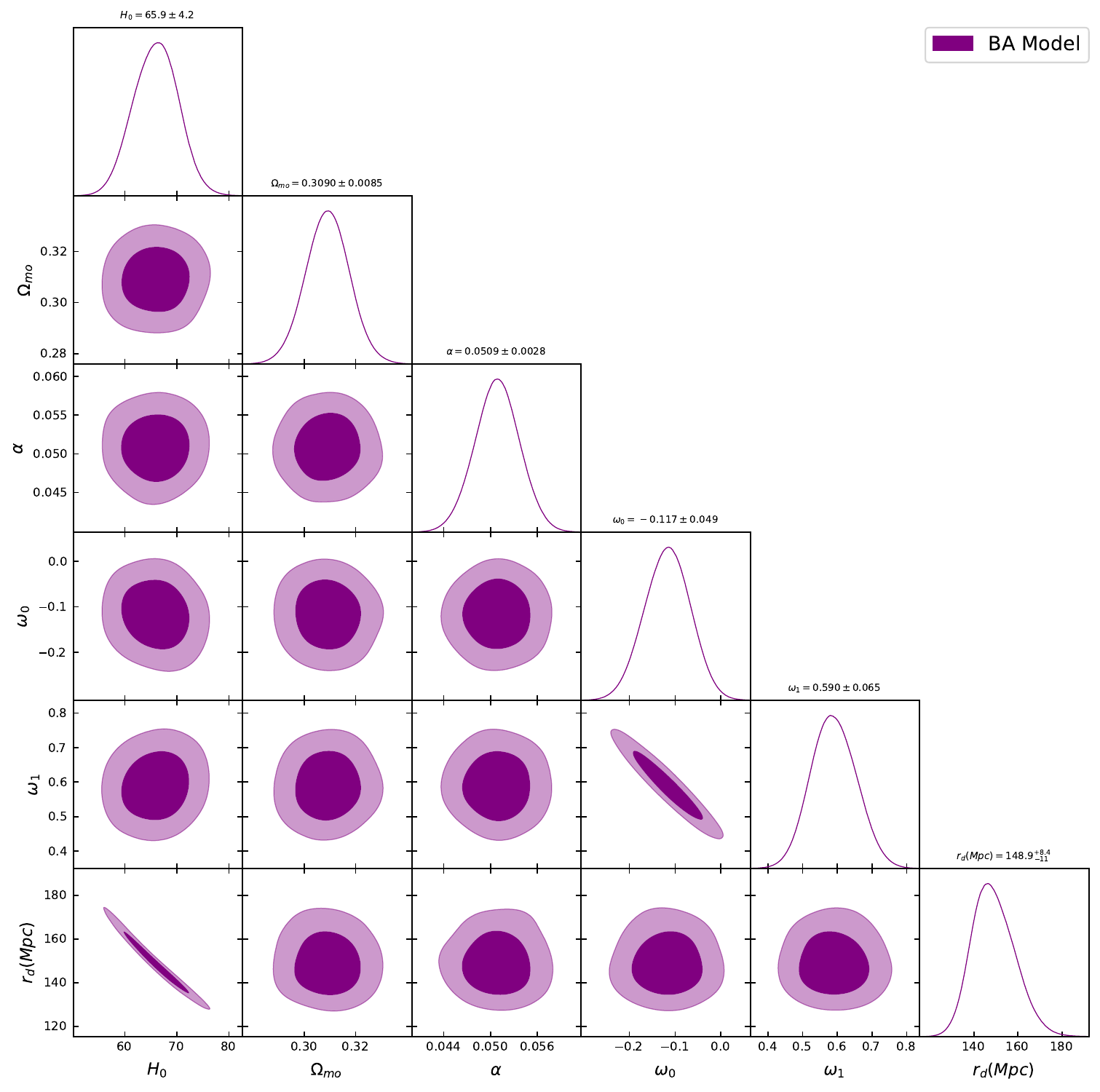}
\end{subfigure}
\caption{Corner plot showing the 1D and 2D marginalized posterior distributions of CPL, JBP, and BA parametrizations within the framework of 4D Einstein Gauss Bonnet gravity at 68\% (1$\sigma$) and 95\% (2$\sigma$) confidence levels}\label{fig_1}
\end{figure*}

\begin{table*}
\begin{tabular}{lcccccccc}
\toprule
\textbf{Model} & $H_0$ [$\mathrm{km\,s^{-1}\,Mpc^{-1}}$] & $\Omega_{m0}$ & $\alpha$ & $\omega_0$ & $\omega_1$ & $r_d$ [Mpc] & $|\Delta \ln \mathcal{Z}_{\Lambda\mathrm{CDM}, \mathrm{Model}}|$ \\
\midrule
\textbf{Flat $\Lambda$CDM} 
& $68.5 \pm 4.6$ & $0.309 \pm 0.0091$ & --- & --- & ---  & $147.0_{-11}^{+8.8}$ & 0 \\
\addlinespace[0.2cm]
\textbf{DE Parametrization } & & & & \\
CPL & $66.0 \pm 4.3$ & $0.309 \pm 0.0086$ & $0.0509 \pm 0.0029$ & $-0.169 \pm 0.054$ & $1.24 \pm 0.14$ & $149.0_{-11}^{+8.2}$ & 12.71 \\
JBP & $65.1 \pm 4.5$ & $0.309 \pm 0.0086$ & $0.0510 \pm 0.0029$ & $-0.260 \pm 0.083$ & $2.70 \pm 0.39$  & $149.8_{-11}^{+9.6}$ &  2.54 \\
BA & $65.9 \pm 4.2$ & $0.309 \pm 0.0085$ & $0.0509 \pm 0.0028$ & $-0.117 \pm 0.049$ & $0.59 \pm 0.06$  & $148.9_{-11}^{+8.4}$ & 12.74 \\
\bottomrule
\end{tabular}

\caption{Mean values with 68\% (1$\sigma$) credible intervals for the standard $\Lambda$CDM model and CPL, JBP, and BA parametrizations within the framework of 4D Einstein Gauss Bonnet gravity}
\label{tab_1}
\end{table*}
\subsection{Observational and Statistical Results}
Fig~\ref{fig_1} shows the corner plot that shows the joint and marginal posterior distributions of the CPL, JBP , and BA parametrizations 4D Einstein Gauss Bonnet gravity theory. The diagonal elements of the plot represent the 1D marginalized posterior distributions for each cosmological parameter. The off-diagonal elements show the 2D joint posterior distributions between pairs of parameters. These contour plots illustrate the correlations between parameters, with the contours representing the 68\% and 95\% confidence levels.

Table~\ref{tab_1} shows the mean values of various cosmological parameters along with their 68\%  1$\sigma$ credible intervals. It can be observed that the predicted values of $H_0$, $\Omega_{m0}$, and $r_d$ are consistent within the expected ranges. In our analysis, we compare the predicted values of $H_0$, $\Omega_{m0}$, and $r_d$ for the CPL, JBP, and BA models in the framework of 4D Einstein–Gauss–Bonnet gravity with those of the standard $\Lambda$CDM model. Our results show mild deviations in the Hubble parameter, with tensions of approximately 0.4$\sigma$, 0.53$\sigma$, and 0.42$\sigma$ for the CPL, JBP, and BA models, respectively, relative to $\Lambda$CDM. For $\Omega_{m0}$, all models predict values consistent with the $\Lambda$CDM model. Furthermore, the predicted values of the sound horizon $r_d$ for the CPL, JBP, and BA models show mild deviations relative to the $\Lambda$CDM model. Specifically, the tensions in $r_d$ are approximately 0.15$\sigma$, 0.20$\sigma$, and 0.14$\sigma$ for the CPL, JBP, and BA models, respectively.

To compare each model statistically, we compute the Bayes factor ($\ln B_{ab} \equiv \Delta \ln \mathcal{Z}$) relative to the $\Lambda$CDM model. Based on the revised Jeffreys scale \cite{kass1995bayes}, we find that the CPL and BA models exhibit decisive evidence in favor of these parametrizations, with $\Delta \ln \mathcal{Z}$ values of 12.71 and 12.74, respectively. In contrast, the JBP model shows only moderate evidence ($\Delta \ln \mathcal{Z} = 2.54$) compared to $\Lambda$CDM. These results indicate that while all models remain broadly consistent with $\Lambda$CDM in their parameter estimates, the CPL and BA parametrizations are statistically preferred in the context of 4D Einstein Gauss Bonnet gravity.

\section{Mass accretion phenomena of compact objects in 4-dimensional Einstein-Gauss-Bonnet (EGB) gravity theory:}\label{mass_accretion}
To analyze 4D EGB gravity from a border perspective, we now discuss the mass accretion concept of the black hole and the wormhole in succession. Mass accretion is a legitimate way to unify the modified gravity effects and serve as a tool for testing deviations from the usual general relativity in the case of black hole physics, gravitational waves, and other cosmic scenarios. In fact, through future observation of gravitational wave signals, accretion disks, and cosmological structures, we can constrain the Gauss-Bonnet coupling parameter~$\alpha$~which in turn can validate or falsify the 4D EGB gravity theory as an alternative to the general relativity theory. This process is crucial to understanding the observable features and dynamics of black holes and wormholes. Through mass accretion phenomena of black holes, we can investigate their growth over time, hinting at the cosmic evolution of the universe, as well as the energy emissions
that produce the much-needed evidence to detect them. Also, this process helps to observe the interaction between black holes and their surrounding galaxies. Besides, through the accretion phenomena of wormholes, we can predict the stability and traversability properties of wormholes. Briefly, we can say that the mass accretion phenomenon is a potent tool for understanding black hole astrophysics and a key concept for observing the speculative physics of wormholes. So, in this section, let us discuss the mass accretion process of black holes and wormholes in the 4D EGB gravity scenario.\\\\
Considering that dark energy is accreting in the form of a perfect fluid with~$p$~as its pressure and~$\rho$~as its energy density, the energy-momentum tensor can be written as follows:
\begin{equation}\label{energy_momentum_tensor}
T_{_{\mu\nu}}=(\rho+p)u_{_{\mu}}u_{_{\nu}}+pg_{_{\mu\nu}},   
\end{equation}
where,~$u=\frac{dx^{\mu}}{ds}$~represents the fluid four velocity with the normalize condition~$u^{{\mu}}u_{_{\mu}}=-1$.
In terms of fluid-like dark energy accretion, the mass of the black hole is always dynamic, hence it depends on the time variable~$t$. So, for calculating the rate of change in the mass of the 4D EGB black hole, we need to integrate the flux of the fluid over the 4-dimensional volume of the black hole, which gives the following form of the equation of rate of change in the mass of the black hole:
\begin{equation}\label{mass_rate_BH}
 \dot{M}= 4\pi AM^2(\rho+p)~.
\end{equation}
Here,~$M$~is the black hole mass and~$A$~is a positive energy-flux related constant  \cite{babichev2004black,babichev2005accretion,babichev2013black,mukherjee2024accretionphenomenadifferentkinds}. From the above equation, it is obvious that the mass of the black hole increases when the term~$\rho+p$~is positive, that is basically the dark energy case. The possible physical interpretation behind this mass increase is that dark energy generally consists of a potential and a scalar field, and the positive kinetic energy of the scalar field particles present in dark energy can increase the mass of the black hole. In a similar manner, when the term~$\rho+p$~is negative, the black hole mass decreases, marking the phantom energy case. The reason may be that the particle present in the accreting phantom energy has a total negative kinetic energy. In fact, the same concept applies in the case of the Hawking radiation process, where a similar type of particle with negative kinetic energy arises and takes part in the Penrose model of black hole rotational energy extraction. As for the $\Lambda$CDM model of the universe, no significant change in black hole mass due to accretion can be observed, as no kinetic term is involved here. That is why the $\Lambda$CDM model is called the transition state between phantom and non-phantom models of dark energy.\\\\
So, following the conservation equation,~Eq.(\ref{conservation.equ}),~we have the final form of the black hole mass equation in the environment of 4D EGB gravity as follows:
\begin{equation}\label{BH_Mass_equ}
M = \frac{M_0}{1 + \frac{4\pi A M_0}{3} \displaystyle \int_{\rho_0}^{\rho} \frac{1}{H(z)} \, d\rho}~~~,
\end{equation}
where \( H(z) \) is given by Eq.(\ref{Hubble_parameter.equation}), \( M_0 \) represents the mass of the 4D EGB black hole, and \( \rho_0 \) is the total energy density of the accreting dark energy fluid at the present epoch, which can be written as: $\rho_{_{0}}=\rho_{_{DM0}}+\rho_{_{DE0}}$.\\\\
Furthermore, the 4D EGB wormhole also shows a dynamic nature as a result of the accretion of fluid in it. If we take $\mathbb{M}$~as the exotic mass of the wormhole, then it can be considered spherically distributed along the throat of the wormhole \cite{gonzalez2006some}. In fact, in this case, the energy density, pressure, and related integration constants are all time-dependent. So, according to the energy conservation law, the exotic mass and the throat radius of the wormhole become dynamic because of the accreting dark energy fluid. It shows that even if the wormhole metric is static in the beginning, the stored energy in it must be dynamic due to a dark energy type of fluid accretion. Following \cite{gonzalez2006some}~, we can analyze the effect of dark energy accretion on the wormhole simply by calculating the general rate of change in the stored energy of the wormhole due to external accretion. The rate of change in the exotic mass of the wormhole following the momentum density is given as~$ \mathbb{\dot{M}}=\int T_0^1~ds$. This concept is inspired by the accretion phenomena of higher-dimensional wormholes \cite{debnath2020nature}.\\\\
Hence, integrating the flux of the accreting fluid over the 4-dimensional volume of the wormhole, we can formulate the expression of the rate of change of the mass function as given below \cite{gonzalez2006some,debnath2020nature}:
\begin{equation}\label{mass_rate_WH}
\mathbb{\dot{M}}=-4\pi B\mathbb{M}^2(\rho+p)~,    
\end{equation}
where~$ B $~is nothing but a positive constant. From the above analysis, we can clearly observe that asymptotically the rate of change in the exotic mass of the 4D EGB wormhole due to dark energy accretion is exactly the negative of the corresponding rate for the 4D EGB black hole~(see Eq.(\ref{mass_rate_BH})~and~Eq.(\ref{mass_rate_WH})). As the 4D EGB wormhole is static in nature, its mass depends only on the radial coordinate~$r$. Thus, accretion of any type of fluid around the wormhole makes the wormhole mass~$ \mathbb{M} $~a dynamical function of the time variable~$ t $~. It establishes the fact that $\mathbb{\dot{M}}$~depends on time and therefore the exotic mass of the 4D EGB wormhole~$\mathbb{M}$~has a sensitive dependency on the nature of the accreting fluid. This dependence is shown through the specific term $\rho+p$. So, for the phantom-like dark energy fluid~($\rho+p<0$), the mass of the wormhole increases, while for the non-phantom dark energy fluid ~($\rho+p>0$), the wormhole mass decreases \cite{gonzalez2006some}.\\[1.5mm]
The final form of the exotic mass equation for the 4D EGB wormhole follows from the conservation equation,~Eq.(\ref{conservation.equ}),~given as below:
\begin{equation}\label{WH_Mass_equ}
\mathbb{M}=\frac{\mathbb{M}_0}{1-\frac{4\pi B\mathbb{M}_0}{3} \displaystyle\int_{\rho_0}^{\rho} \frac{1}{H(z)} d\rho}~~~. 
\end{equation}     
Where \( H(z) \) is given by Eq.(\ref{Hubble_parameter.equation}) and ~$\mathbb{M}_0$~represents the mass of the 4D EGB wormhole in the present epoch.\\\\
As we already mentioned, the changes in the 4D EGB black hole mass and the 4D EGB wormhole mass sensitively depend on the nature of the accreting fluid. So, it is of great importance to consider ideal candidates for this part. In this work, we consider some mainstream redshift parameterizations of the equation of state parameter~(see Sec.\ref{CPL},~Sec.\ref{JBP},~Sec.\ref{BA})~rather than simply considering some dark energies, which can make this work more relevant from the accuracy and significance perspective. The parameterization forms of dark energy are more data-driven and give a better understanding of the evolution of the universe. Here, let us discuss how these dark-energy parameterizations can influence the mass accretion phenomena of the black hole and wormhole in the context of 4D EGB gravity.

\subsection{Accretion of Dark Energy with Chevallier-Polarski-Linder (CPL) Parametrization:}\label{mass_accretion_CPL}
\begin{figure*}
\begin{subfigure}{0.42\textwidth}
\includegraphics[width=\linewidth]{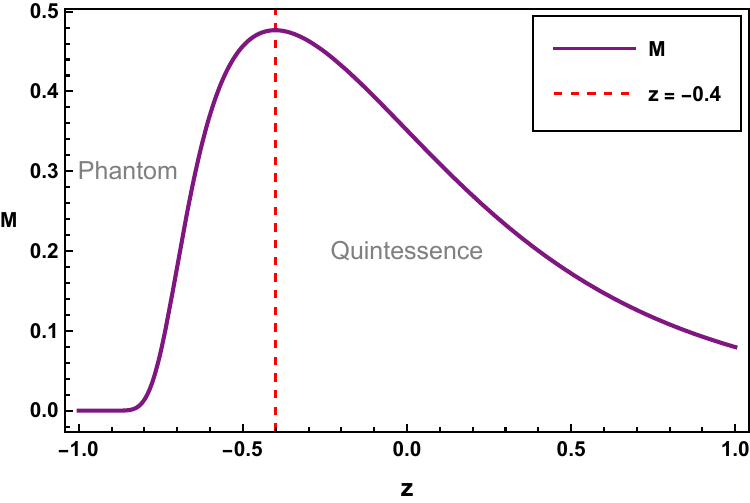}
    \subcaption{The role of CPL parameterization type dark energy on the mass accretion process of a 4D EGB black hole }
    \label{fig:f 2~(a)}
\end{subfigure}
\hfill
\begin{subfigure}{0.42\textwidth}
\includegraphics[width=\linewidth]{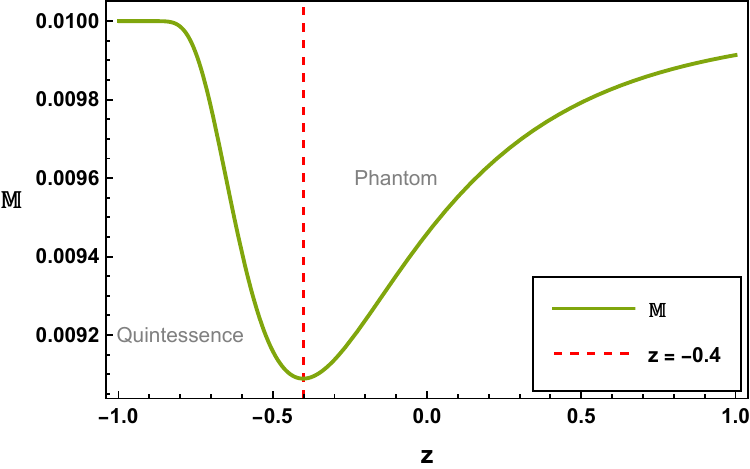}
    \subcaption{The role of CPL parameterization type dark energy on the mass accretion process of a 4D EGB wormhole }
    \label{fig:f 2~(b)}
\end{subfigure}
\caption{Effects of CPL-type dark energy parameterization model in the mass accretion process of compact objects in the 4D EGB gravity background }
\label{fig:my_label2} 
\end{figure*}

The mass function of the 4D EGB black hole due to the CPL parameterization type dark energy accretion is given as follows:
\begin{equation}\label{BH_Mass_CPL}
M=\frac{M_0}{1+\frac{4\pi AM_0}{3} \displaystyle\int_{\rho_{_{_{TCPL0}}}}^{\rho_{_{_{TCPL}}}} \frac{1}{H_{_{CPL}} (z)}    d\rho_{_{TCPL}}}~~~. 
\end{equation}
Here, \( H_{_{CPL}}(z) \) is given by Eq.(\ref{Hubble_parameter_CPL}), and \( \rho_{_{_{TCPL}}} \) represents the total energy density due to the accretion of dark energy of the CPL parameterization type, which can be written using  Eq.(\ref{energy_density_equation.DM})~,~Eq.(\ref{DE_relation})~and~Eq.(\ref{energy_density_CPL}) together as given below:
\begin{equation}\label{Total_energy_density_CPL}
\begin{split}
\rho_{_{_{TCPL}}} = & \frac{3H_0^2}{8\pi G} \left[ \Omega_{_{m0}} (1+z)^3 + \right. \\
  & \left. \left( 1 + \alpha H_0^2 - \Omega_{_{m0}} \right) \left( 1+z \right)^{3\left(1+\omega_0+\omega_1\right)} \right. \\
  & \left. \exp\left( \frac{-3\omega_1 z}{1+z} \right) \right],
\end{split}
\end{equation}
and ~$\rho_{_{_{TCPL0}}}$~is the present value of the total energy density.\\
Again, the exotic mass of the 4D EGB wormhole due to the CPL parameterization-type dark energy accretion is given in the following form:
\begin{equation}\label{WH_Mass_CPL}
\mathbb{M}=\frac{\mathbb{M}_0}{1-\frac{4\pi B\mathbb{M}_0}{3} \displaystyle\int_{\rho_{_{_{TCPL0}}}}^{\rho_{_{TCPL}}} \frac{1}{ H_{_{_{CPL}}}(z)} d\rho_{_{_{TCPL}}}}~~~.  \end{equation}   
Now we can visualize the effects this CPL parameterization type of dark-energy accretion creates on the mass of the black hole and wormhole in the framework of 4D EGB gravity during the evolution of the universe through a mass vs. redshift graph represented by Fig.\ref{fig:my_label2}.
\subsection{Accretion of Dark Energy with Jassal-Bagla-Padmanabhan (JBP) Parametrization:}\label{mass_accretion_JBP}
\begin{figure*}
\begin{subfigure}{0.42\textwidth}
\includegraphics[width=\linewidth]{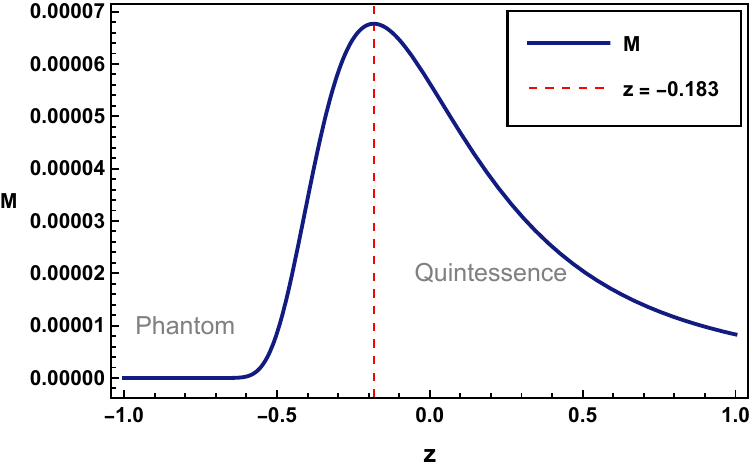}
    \subcaption{The role of JBP parameterization type dark energy on the mass accretion process of a 4D EGB black hole }
    \label{fig:f 3~(a)}
\end{subfigure}
\hfill
\begin{subfigure}{0.42\textwidth}
\includegraphics[width=\linewidth]{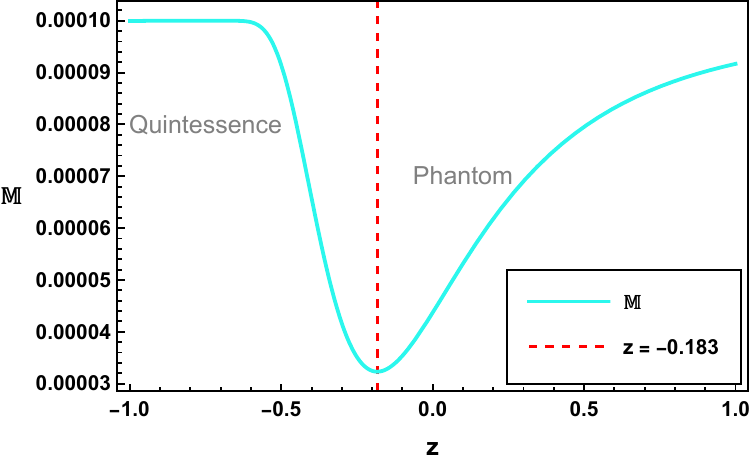}
    \subcaption{The role of JBP parameterization type dark energy on the mass accretion process of a 4D EGB wormhole }
    \label{fig:f 3~(b)}
\end{subfigure}
\caption{Effects of JBP-type dark energy parameterization model in the mass accretion process of compact objects in the 4D EGB gravity background }
\label{fig:my_label3} 
\end{figure*}

The mass function of the 4D EGB black hole due to the JBP parameterization type dark energy accretion is given as follows:
\begin{equation}\label{BH_Mass_JBP}
M=\frac{M_0}{1+\frac{4\pi AM_0}{3} \displaystyle\int_{\rho_{_{_{TJBP0}}}}^{\rho_{_{_{TJBP}}}} \frac{1}{ H_{_{_{JBP}}}(z)} d\rho_{_{_{TJBP}}}}~~~,     
\end{equation}    
Here, \( H_{_{_{JBP}}}(z) \) is given by Eq.(\ref{Hubble_parameter_JBP}), and $\rho_{_{_{TJBP}}}$~is the total energy density due to the accretion of the dark energy of the JBP parameterization type, which can be formulated using Eq.(\ref{energy_density_equation.DM})~,~Eq.(\ref{DE_relation})~and~Eq.(\ref{energy_density_JBP}) together as follows:
\begin{equation}\label{Total_energy_density_JBP}
\begin{split}
\rho_{_{_{TJBP}}} = & \frac{3H_0^2}{8\pi G} \left[ \Omega_{_{m0}} (1+z)^3 + \right. \\
  & \left. \left( 1 + \alpha H_0^2 - \Omega_{_{m0}} \right) \left( 1+z \right)^{3(1+\omega_0)} \right. \\
  & \left. \exp\left( \frac{3\omega_1 z^2}{2(1+z)^2} \right) \right],
\end{split}
\end{equation}
and ~$\rho_{_{_{TJBP0}}}$~is the present value of the total energy density.\\
Again, the exotic mass of the 4D EGB wormhole due to the JBP parameterization-type dark energy accretion is given in the following form:
 \begin{equation}\label{WH_Mass_JBP}
\mathbb{M}=\frac{\mathbb{M}_0}{1-\frac{4\pi B\mathbb{M}_0}{3} \displaystyle\int_{\rho_{_{_{TJBP0}}}}^{\rho_{_{_{TJBP}}}} \frac{1}{H_{_{JBP}}(z)} d\rho_{_{_{TJBP}}}}~~~.     
 \end{equation}
Hence, we can visualize the effects this JBP parameterization-type dark energy accretion creates on the mass of the black hole and wormhole in the framework of 4D EGB gravity during the evolution of the universe through a mass vs. redshift graph represented by Fig.\ref{fig:my_label3}.
\subsection{Accretion of Dark Energy with Barboza-Alcaniz (BA) Parametrization:}\label{mass_accretion_BA}
The mass function of the 4D EGB black hole due to the BA parameterization type dark energy accretion is given as follows:
\begin{equation}\label{BH_Mass_BA}
M=\frac{M_0}{1+\frac{4\pi AM_0}{3} \displaystyle\int_{\rho_{_{_{TBA0}}}}^{\rho_{_{_{TBA}}}} \frac{1}{ H_{_{BA}}(z)} d\rho_{_{_{TBA}}}}~~~,     
\end{equation}
Here \( H_{_{BA}}(z) \) is given by Eq.(\ref{Hubble_parameter_BA}), and $\rho_{_{_{TBA}}}$~is the total energy density due to the accretion of the dark energy of the BA parameterization type, which can be formulated using Eq.(\ref{energy_density_equation.DM})~,~Eq.(\ref{DE_relation})~and~Eq.(\ref{energy_density_BA}) together as follows:
\begin{equation}\label{Total_energy_density_BA}
\begin{split}
\rho_{_{_{TBA}}} = & \frac{3H_0^2}{8\pi G} \left[ \Omega_{_{m0}} (1+z)^3 + \right. \\
  & \left. \left( 1 + \alpha H_0^2 - \Omega_{_{m0}} \right) \left( 1+z \right)^{3(1+\omega_0)} \right. \\
  & \left. \left( 1+z^2 \right)^{\frac{3\omega_1}{2}} \right],
\end{split}
\end{equation}
and ~$\rho_{_{_{TBA0}}}$~is the present value of the total energy density.\\

\begin{figure*}
\begin{subfigure}{0.42\textwidth}
\includegraphics[width=\linewidth]{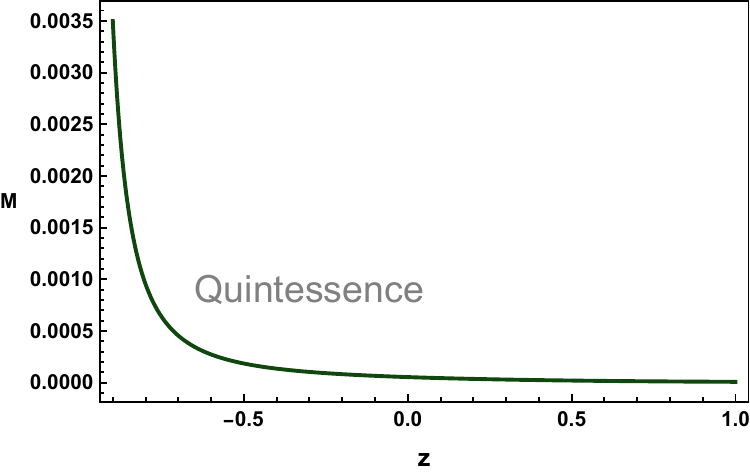}
    \subcaption{The role of BA parameterization type dark energy on the mass accretion process of a 4D EGB black hole}
    \label{fig:f 4~(a)}
\end{subfigure}
\hfill
\begin{subfigure}{0.42\textwidth}
\includegraphics[width=\linewidth]{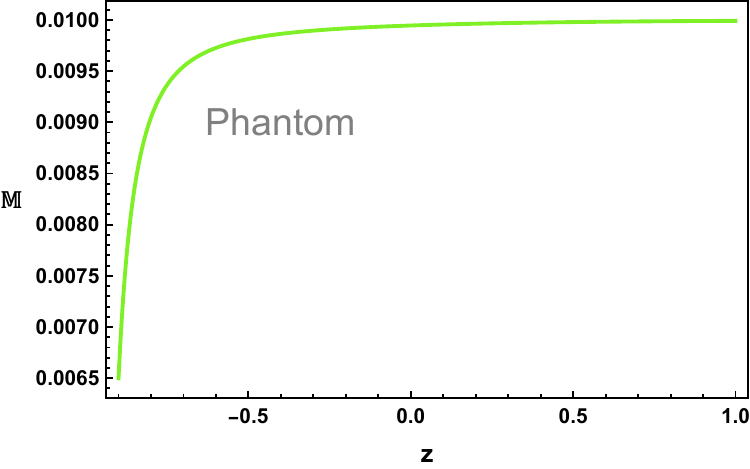}
    \subcaption{The role of BA parameterization type dark energy on the mass accretion process of a 4D EGB wormhole }
    \label{fig:f 4~(b)}
\end{subfigure}
\caption{Effects of BA-type dark energy parameterization model in the mass accretion process of compact objects in the 4D EGB gravity background }
\label{fig:my_label4} 
\end{figure*}

\begin{figure*}
\begin{subfigure}{0.42\textwidth}
\includegraphics[width=\linewidth]{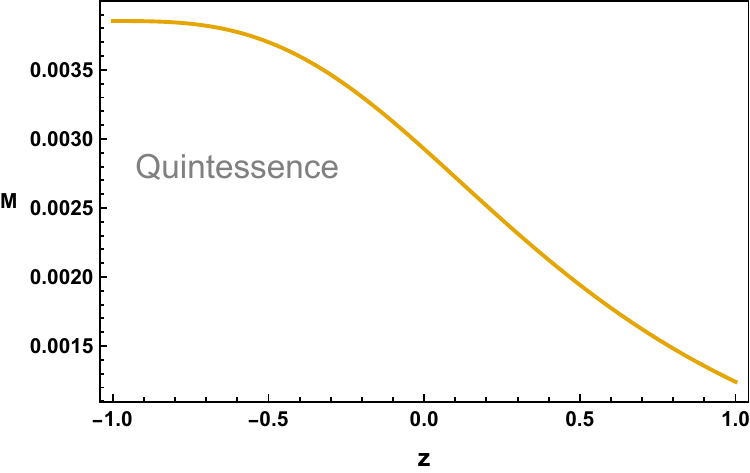}
    \subcaption{Effects of the standard $\Lambda$CDM model on the mass of a 4D EGB black hole during the evolution of the universe }
    \label{fig:f 5~(a)}
\end{subfigure}
\hfill
\begin{subfigure}{0.42\textwidth}
\includegraphics[width=\linewidth]{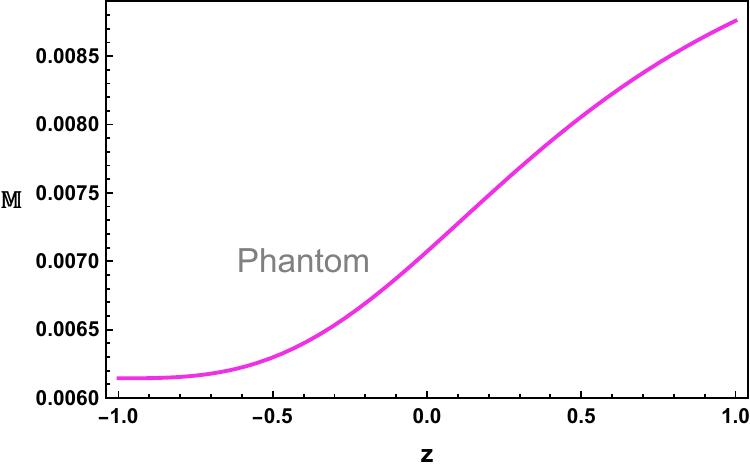}
    \subcaption{Effects of the standard $\Lambda$CDM model on the mass of a 4D EGB wormhole during the evolution of the universe }
    \label{fig:f 5~(b)}
\end{subfigure}
\caption{The role of the standard $\Lambda$CDM model in the mass accretion process of compact objects in the 4D EGB gravity background }
\label{fig:my_label5} 
\end{figure*}

Again, the exotic mass of the 4D EGB wormhole due to the BA parameterization type dark energy accretion is given by the following form:
 \begin{equation}\label{WH_Mass_BA}
\mathbb{M}=\frac{\mathbb{M}_0}{1-\frac{4\pi B\mathbb{M}_0}{3} \displaystyle\int_{\rho_{_{_{TBA0}}}}^{\rho_{_{_{TBA}}}} \frac{1}{ H_{_{BA}}(z)}d\rho_{_{_{TBA}}}}~~~.     
 \end{equation}   
Thus, we can visualize the effects this BA parameterization type of dark-energy accretion creates on the mass of the black hole and wormhole in the framework of 4D EGB gravity during the evolution of the universe through a mass vs. redshift graph represented by Fig.\ref{fig:my_label4}.\\\\
Furthermore, we also graphically illustrate the variations in the mass function of a 4D EGB black hole and a 4D EGB wormhole, respectively, with changes in the redshift function, considering the effects of the standard~$\Lambda$CDM model of the Universe in Fig.\ref{fig:my_label5}. 
\subsection{Interpretation:}\label{interpre}
Let us now discuss the results of the mass accretion phenomena of compact objects in the 4D EGB gravity-induced universe, portrayed through the mass vs. redshift graphs for all three different parameterization cases as well as the standard $\Lambda$CDM case.
\begin{itemize}
    \item In the Fig.\ref{fig:my_label2}~the mass vs. redshift graphs are drawn for a 4D EGB black hole and a 4D EGB wormhole side by side, showing the changes in their mass during the evolution of the universe for the accretion of CPL-type dark energy parameterization. This graph shows a combination of the two most important and completely opposite eras of the universe, the quintessence era for the redshift limit~$ z>-0.4$~~where the mass of the black hole increases gradually, contributing to the possible expansion of the universe, which exhibits its nature as dark energy, but for the redshift value~$z<-0.4$~it shows the nature of the phantom energy by gradually reducing the black hole mass with the evolution of the universe. The line ~$z=-0.4$~(shown in a red dashed line in Fig.~\ref{fig:f 2~(a)}) works as a separator for these two stages. It also shows how the accretion of the same dark energy parameterization model can lead to two completely different fates of the black hole mass, depending solely on the different redshift limits. So, accretion of CPL-type dark energy parameterization first gradually increases the mass of a 4D EGB black hole up to a certain redshift value~($z=-0.4$), and then it starts to gradually decrease the mass.\\
    On the contrary, in the wormhole case, the line ~$z=-0.4$~(shown in a red dashed line in Fig.~\ref{fig:f 2~(b)}) works as a partition for the phantom ($z>-0.4$)~stage and quintessence ($z<-0.4$)~stage. So, accretion of CPL-type dark energy parameterization first gradually decreases the mass of a 4D EGB wormhole up to a certain redshift value~($z=-0.4$), and then it starts to gradually increase the mass, marking the exact opposite of the 4D EGB black hole case. It also demonstrates the evolving behavior of the CPL parameterization type of dark energy.

    \item In the Fig.\ref{fig:my_label3}~the mass vs. redshift graphs are drawn for the accretion of JBP-type dark energy parameterization. This graph again shows two completely different eras of the universe, the quintessence era for the redshift limit~$ z>-0.183$~~where the mass of the black hole increases gradually, contributing to the possible expansion of the universe, which exhibits its nature as dark energy, but for the redshift value~$z<-0.183$~it shows the nature of the phantom energy by gradually reducing the black hole mass with the evolution of the universe. Also, the line ~$z=-0.183$~(shown in a red dashed line in Fig.~\ref{fig:f 3~(a)}) works as a separator for these two stages. So, we can conclude that the accretion of JBP-type dark-energy parameterization first gradually increases the mass of a 4D EGB black hole up to a certain redshift value~($z=-0.183$), and then it starts to gradually decrease the mass.\\
    Contrarily, for the wormhole case, the line ~$z=-0.183$~(shown in a red dashed line in Fig.~\ref{fig:f 3~(b)}) works as a partition for the phantom ($z>-0.183$)~stage and quintessence ($z<-0.183$)~stage. So, accretion of JBP-type dark energy parameterization first gradually decreases the mass of a 4D EGB wormhole up to a certain redshift value~($z=-0.183$), and then it starts to gradually increase the mass, resembling a striking opposition of the 4D EGB black hole nature. Thus, the JBP type of dark energy parameterization also shows an evolving nature.
    
    \item In the Fig.\ref{fig:my_label4}~the mass vs. redshift graphs are drawn for the accretion of BA-type dark energy parameterization. In this case, the graph only shows the increasing pattern of the mass function for the 4D EGB black hole case~(see fig.\ref{fig:f 4~(a)}), indicating the quintessence era of the universe, which reflects the nature of a true dark energy and contributes to the possible expansion rate of the universe.\\
    While in the case of 4D EGB wormhole~(see Fig.\ref{fig:f 4~(b)}),~as predicted, it shows the exact opposite scenario compared to the 4D EGB black hole case. In this case, the graph only shows the decreasing pattern of the mass function, indicating the nature of a phantom-like dark energy.
   
    \item Finally, in the Fig.\ref{fig:my_label5}, we picturised the mass vs. redshift graphs of the black hole and wormhole side by side for the standard $\Lambda$CDM model in the framework of the 4D EGB gravity. It shows an increasing trait for the black hole case, interpreting the quintessence era of the universe~(see Fig.\ref{fig:f 5~(a)})~, whereas a decreasing trait for the wormhole case~(see Fig.\ref{fig:f 5~(b)})~, interpreting the phantom era of the universe, which is in accordance with other previous theoretical findings related to this topic. 
   \end{itemize}
All these graphical representations are constructed using parametric values, which are constrained through the MCMC technique given in the Table.\ref{tab_1}~of Sec.\ref{Data}~for more reliable, accurate and physically plausible results. 


\section{Discussion on the results and concluding remarks:}\label{conclusion}
The modified theory of gravity is one of the most remarkable developments in modern cosmology. Beyond its capacity to provide new testable physics that can exceed general relativity, it can also resolve its theoretical limitations. Although such theories cannot yet replace Einstein's general relativity framework, they can still provide us with a vital way to navigate the new domain of gravitational science. As we mentioned earlier, 4-dimensional Einstein-Gauss-Bonnet (EGB) gravity is one of such modified gravity theories that can connect quantum theory, string theory, and the theory of gravitation altogether, making it an emerging topic of current research. In this work, we studied this particular gravity theory from the context of the mass accretion process of compact objects, such as black holes and wormholes. Although by virtue of the modified gravity theory,~4D EGB gravity can itself initiate dark energy, in this work, we consider some other dark energy parameterization models, excluding this hypothesis. The summary of our work is listed below. 
\begin{itemize} 
\item In Sec.\ref {EGB_thory}, we first discuss the motivation and basic concepts behind the 4D EGB gravity theory. Then, one by one, we consider the associated modified Friedmann equations and discuss the black hole metric, as well as the wormhole metric, in the framework of 4D EGB gravity. With the assumption that the background universe only consists of dark matter and dark energy, we obtain the energy density equation for dark matter from the separate application of the continuity equation. Next, we deduce the corresponding expression of the Hubble parameter in terms of the redshift function $z$~for the 4D EGB gravity background, and a specific relationship between dark energy, dark matter, and the Gauss-Bonnet coupling parameter.

\item In Sec.\ref {DE_parametrization}, we explain the concept of dark energy parameterization and give a detailed insight about our chosen three parameterizations, namely the Chevallier-Polarski-Linder (CPL) parameterization, the Jassal-Bagla-Padmanabhan (JBP) parameterization, and the Barboza-Alcaniz (BA) parameterization. First, we write down their corresponding expressions for the equation of state parameter, together with their energy density equations. Then, one by one, we obtain the Hubble parameter equation for each model in terms of the redshift function $z$. 

\item  In Sec.\ref {Data}, we study the comparisons of three dark energy equation of state models in the context of 4D Einstein–Gauss–Bonnet gravity theory: CPL, JBP, and BA, alongside the \(\Lambda\)CDM model. The values of \(H_0\) and \(r_d\) are consistent with the Planck 2018 results for all models. The matter density parameter, \(\Omega_{m0}\), is nearly identical for the \(\Lambda\)CDM and CPL models (\(0.3126 \pm 0.0055\)), while JBP and BA predict slightly higher values (\(0.3170 \pm 0.0057\) and \(0.3169 \pm 0.0055\), respectively). The CPL model shows a moderate deviation from \(\omega_0 = -1\) (\(\omega_0 = -0.245 \pm 0.054\) and \(\omega_1 = 1.43 \pm 0.14\)), while JBP predicts a stronger evolution with \(\omega_0 = -0.437 \pm 0.080\) and \(\omega_1 = 3.54 \pm 0.38\). The BA model provides milder deviations (\(\omega_0 = -0.183 \pm 0.048\) and \(\omega_1 = 0.68 \pm 0.06\)). When compared with DESI DR2, which reports a more negative \(w_0\) (\(-1.23_{-0.61}^{+0.44}\)) and limited evolution of dark energy, the 4D Einstein Gauss-Bonnet models, particularly JBP, predict stronger evolution. Non-CMB studies give \(w_0 = -0.876 \pm 0.055\) and \(w_a = 0.10_{+0.32}^{-0.20}\), which are closer to the CPL and BA models. Bayesian analysis shows strong evidence for the CPL and BA models over \(\Lambda\)CDM, with Bayes factors of \( \ln(B_{\text{CPL-LCDM}}) = 9.3903 \) and \( \ln(B_{\text{BA-LCDM}}) = 9.4878 \), while the JBP model has a Bayes factor of \( \ln(B_{\text{JBP-LCDM}}) = -3.0154 \), suggesting strong evidence against it in favor of \(\Lambda\)CDM. Overall, the CPL and BA models are favored, with the JBP model being less favorable.

\item In Sec.\ref{mass_accretion}, we first describe the importance of the mass accretion phenomena regarding modified gravity theories, specifically our considered 4D EGB gravity model, as well as in analyzing black hole and wormhole physics. Next, we investigate the mass accretion process of the 4D EGB black hole and 4D EGB wormhole in a respective manner. Firstly, we establish the relationship between the black hole mass function~$ M$~and the redshift function~$z$~and the relation of wormhole mass function~$\mathbb{M}$~ with the redshift function~$z$~respectively, considering the accreting dark energy in the form of a perfect fluid. Finally, we consider the chosen three dark-energy parameterization models one by one to observe their effects on the black hole and wormhole masses in the background of 4D EGB gravity. Additionally, we examine the changes in the mass pattern of the 4D EGB black hole and 4D EGB wormhole within the standard $\Lambda$CDM model of the universe. All of these results are represented in the form of mass vs. redshift plots~(see Fig.\ref{fig:my_label2}~to~Fig.\ref{fig:my_label5})~derived using the constrained values of parameters, presented in Table.\ref{tab_1} of the previous section. Finally, we explain all the findings of this mass accretion process thoroughly in Sec.\ref{interpre}. 
\end{itemize}
So, to summarize this study, we can conclude that after the accretion of CPL and JBP parameterization-type dark energies, the mass of a 4D EGB black hole first gradually increases and then gradually decreases, showing a transition from quintessence to phantom era during the evolution of the universe, which demonstrates the dynamic nature of these dark energies. But the accretion of BA parameterization-type dark energy can only increase the 4D EGB black hole mass, thereby indicating the quintessence phase of the universe. Conversely, in the case of a 4D EGB wormhole for the accretion of CPL and JBP parameterization-type dark energies, the transition is from a phantom to a quintessence era as the mass first gradually decreases and then gradually increases, again indicating their evolving nature. Also, the accretion of BA parameterization type dark energy can only decrease the 4D EGB wormhole mass, thus showing the phantom phase of the universe. We can also conclude from this study that both the CPL and the JBP parameterization models can behave sometimes as dark energy and sometimes as phantom energy, depending entirely on the values of the redshift function~$z$, revealing the changing properties of dark energy with the evolution of the universe, while the BA parameterization model shows only the nature of dark energy for the 4D EGB gravity-induced universe. As previously predicted in various theoretical findings, our results also verify that the nature of both black holes and wormholes is completely opposite during the evolution of the universe, which is also supported by constrained parameter values.\\ 
    
\section*{ACKNOWLEDGEMENTS:}
PM appreciates IIEST, Shibpur, India, for providing an Institute Fellowship (SRF).\\\\

\section*{Data Availability Statement}
All datasets used in this study will be made available upon reasonable request. This includes the cosmological datasets, numerical outputs, and code.

\bibliographystyle{elsarticle-num}
\bibliography{reference.bib}

@article{kumar2020hayward,
  title={\href{https://doi.org/10.1016/j.aop.2020.168214}{Hayward black holes in the novel 4D Einstein-Gauss-Bonnet gravity}},
  author={Kumar, Arun and Ghosh, Sushant G},
  journal={arXiv preprint arXiv:2004.01131},
  year={2020}
}

@article{kass1995bayes,
  title={Bayes factors},
  author={Kass, Robert E and Raftery, Adrian E},
  journal={Journal of the american statistical association},
  volume={90},
  number={430},
  pages={773--795},
  year={1995},
  publisher={Taylor \& Francis}
}

@article{rayimbaev2022magnetized,
  title={\href{https://doi.org/10.3390/universe8100549}{Magnetized and Magnetically Charged Particles Motion around Regular Bardeen Black Hole in 4D Einstein Gauss-Bonnet Gravity}},
  author={Rayimbaev, Javlon and Bardiev, Dilshodbek and Abdulxamidov, Farrux and Abdujabbarov, Ahmadjon and Ahmedov, Bobomurat},
  journal={Universe},
  volume={8},
  number={10},
  pages={549},
  year={2022},
  publisher={MDPI}
}

@article{goliath2001supernovae,
  title={Supernovae and the nature of the dark energy},
  author={Goliath, M and Amanullah, Rahman and Astier, P and Goobar, A and Pain, R},
  journal={Astronomy \& Astrophysics},
  volume={380},
  number={1},
  pages={6--18},
  year={2001},
  publisher={EDP Sciences}
}

@article{lewis2025getdist,
  title={GetDist: a Python package for analysing Monte Carlo samples},
  author={Lewis, Antony},
  journal={Journal of Cosmology and Astroparticle Physics},
  volume={2025},
  number={08},
  pages={025},
  year={2025},
  publisher={IOP Publishing}
}

@article{handley2015polychord,
  title={PolyChord: nested sampling for cosmology},
  author={Handley, WJ and Hobson, MP and Lasenby, AN},
  journal={Monthly Notices of the Royal Astronomical Society: Letters},
  volume={450},
  number={1},
  pages={L61--L65},
  year={2015},
  publisher={Oxford University Press}
}

@article{sheykhi2023growth,
  title={\href{https://doi.org/10.48550/arXiv.2301.13263}{Growth of perturbations in higher dimensional Gauss-Bonnet FRW cosmology}},
  author={Sheykhi, Ahmad and Farsi, Bita},
  journal={arXiv preprint arXiv:2301.13263},
  year={2023}
}

@article{islam2024strong,
  title={\href{https://doi.org/10.1016/j.cjph.2024.03.044}{Strong gravitational lensing by Bardeen black holes in 4D EGB gravity: constraints from supermassive black holes}},
  author={Islam, Shafqat Ul and Ghosh, Sushant G and Maharaj, Sunil D},
  journal={Chinese Journal of Physics},
  volume={89},
  pages={1710--1724},
  year={2024},
  publisher={Elsevier}
}

@article{haghani2020growth,
  title={\href{https://doi.org/10.1016/j.dark.2020.100720}{Growth of matter density perturbations in 4D Einstein-Gauss-Bonnet gravity}},
  author={Haghani, Zahra},
  journal={Physics of the Dark Universe},
  volume={30},
  pages={100720},
  year={2020},
  publisher={Elsevier}
}

@article{rayimbaev2022shadow,
  title={\href{https://doi.org/10.1142/S0218271822500559}{Shadow and massless particles around regular Bardeen black holes in 4D Einstein Gauss-Bonnet gravity}},
  author={Rayimbaev, Javlon and Bardiev, Dilshodbek and Mirzaev, Temurbek and Abdujabbarov, Ahmadjon and Khalmirzaev, Akram},
  journal={International Journal of Modern Physics D},
  volume={31},
  number={07},
  pages={2250055},
  year={2022},
  publisher={World Scientific}
}

@article{kumar2022bardeen,
  title={\href{https://doi.org/10.3390/universe8040232}{Bardeen Black Holes in the Regularized 4 D Einstein-Gauss-Bonnet Gravity}},
  author={Kumar, Arun and Walia, Rahul Kumar and Ghosh, Sushant G},
  journal={Universe},
  volume={8},
  number={4},
  pages={232},
  year={2022},
  publisher={MDPI}
}

@article{zhang2021bardeen,
  title={\href{https://doi.org/10.48550/arXiv.2112.11869}{Bardeen black hole in magnetically charged four-dimensional Einstein-Gauss-Bonnet gravity}},
  author={Zhang, Shu-Jun and Zhang, He-Xu and Shao, Lei and Deng, Jian-Bo and Hu, Xian-Ru},
  journal={arXiv preprint arXiv:2112.11869},
  year={2021}
}

@article{fernandes20224d,
  title={\href{https://doi.org/10.1088/1361-6382/ac500a}{The 4D Einstein-Gauss-Bonnet theory of gravity: a review}},
  author={Fernandes, Pedro GS and Carrilho, Pedro and Clifton, Timothy and Mulryne, David J},
  journal={Classical and Quantum Gravity},
  volume={39},
  number={6},
  pages={063001},
  year={2022},
  publisher={IOP Publishing}
}

@inproceedings{bardeen1968non,
  title={Non-singular general relativistic gravitational collapse},
  author={Bardeen, James},
  booktitle={Proceedings of the 5th International Conference on Gravitation and the Theory of Relativity},
  pages={87},
  year={1968}
}

@article{boulware1985string,
  title={\href{https://doi.org/10.1103/PhysRevLett.55.2656?_gl=1*k40d97*_ga*MTU0NTU3NTQyNi4xNjk1MjQ4Mjk1*_ga_ZS5V2B2DR1*MTczOTQ1NjA3Ny41MS4wLjE3Mzk0NTYwNzcuMC4wLjgwODEwNTA0OA..}{String-generated gravity models}},
  author={Boulware, David G and Deser, Stanley},
  journal={Physical Review Letters},
  volume={55},
  number={24},
  pages={2656},
  year={1985},
  publisher={APS}
}

@article{mehdizadeh2015einstein,
  title={\href{https://doi.org/10.1103/PhysRevD.91.084004?_gl=1*1c01lwq*_ga*MTU0NTU3NTQyNi4xNjk1MjQ4Mjk1*_ga_ZS5V2B2DR1*MTczOTg4MjMxNy41Mi4wLjE3Mzk4ODIzMTcuMC4wLjE4OTcyNjM2Nzc.}{Einstein-Gauss-Bonnet traversable wormholes satisfying the weak energy condition}},
  author={Mehdizadeh, Mohammad Reza and Zangeneh, Mahdi Kord and Lobo, Francisco SN},
  journal={Physical Review D},
  volume={91},
  number={8},
  pages={084004},
  year={2015},
  publisher={APS}
}

@article{jusufi2020wormholes,
  title={\href{https://doi.org/10.1140/epjc/s10052-020-8287-x}{Wormholes in 4D Einstein--Gauss--Bonnet gravity}},
  author={Jusufi, Kimet and Banerjee, Ayan and Ghosh, Sushant G},
  journal={The European Physical Journal C},
  volume={80},
  number={8},
  pages={698},
  year={2020},
  publisher={Springer}
}

@article{ayon1998regular,
  title={\href{https://doi.org/10.1103/PhysRevLett.80.5056?_gl=1*1r58bjy*_ga*MTU0NTU3NTQyNi4xNjk1MjQ4Mjk1*_ga_ZS5V2B2DR1*MTc0MzI0MzYwNi41My4wLjE3NDMyNDM2MDYuMC4wLjg2MjMzNDI3OQ..}{Regular black hole in general relativity coupled to nonlinear electrodynamics}},
  author={Ayon-Beato, Eloy and Garcia, Alberto},
  journal={Physical review letters},
  volume={80},
  number={23},
  pages={5056},
  year={1998},
  publisher={APS}
}

@article{ayon2005four,
  title={\href{https://doi.org/10.1007/s10714-005-0050-y}{Four-parametric regular black hole solution}},
  author={Ay{\'o}n-Beato, Eloy and Garcia, Alberto},
  journal={General Relativity and Gravitation},
  volume={37},
  pages={635--641},
  year={2005},
  publisher={Springer}
}

@article{bandyopadhyay2021accretions,
  title={\href{https://doi.org/10.1142/S0217732321500814}{Accretions of Tsallis, R{\'e}nyi and Sharma--Mittal dark energies onto higher-dimensional Schwarzschild black hole and Morris--Thorne wormhole}},
  author={Bandyopadhyay, Tanwi and Debnath, Ujjal},
  journal={Modern Physics Letters A},
  volume={36},
  number={12},
  pages={2150081},
  year={2021},
  publisher={World Scientific}
}

@article{biswas2015observational,
  title={\href{https://ui.adsabs.harvard.edu/link_gateway/2015IJTP...54..341B/doi:10.1007/s10773-014-2229-z}{Observational constraints of red-shift parametrization parameters of dark energy in Horava-Lifshitz gravity}},
  author={Biswas, Ritabrata and Debnath, Ujjal},
  journal={International Journal of Theoretical Physics},
  volume={54},
  pages={341--357},
  year={2015},
  publisher={Springer}
}

@article{mehrabi2018growth,
  title={\href{https://doi.org/10.1103/PhysRevD.97.083522}{Growth of perturbations in dark energy parametrization scenarios}},
  author={Mehrabi, Ahmad},
  journal={Physical Review D},
  volume={97},
  number={8},
  pages={083522},
  year={2018},
  publisher={APS}
}

@article{kumar2025exploring,
  title={\href{https://doi.org/10.1103/PhysRevD.111.043503}{Exploring alternative cosmologies with the LSST: Simulated forecasts and current observational constraints}},
  author={Kumar, Dharmendra and Mitra, Ayan and Adil, Shahnawaz A and Sen, Anjan A},
  journal={Physical Review D},
  volume={111},
  number={4},
  pages={043503},
  year={2025},
  publisher={APS}
}

@article{chevallier2001accelerating,
  title={\href{https://doi.org/10.1142/S0218271801000822}{Accelerating universes with scaling dark matter}},
  author={Chevallier, Michel and Polarski, David},
  journal={International Journal of Modern Physics D},
  volume={10},
  number={02},
  pages={213--223},
  year={2001},
  publisher={World Scientific}
}

@article{linder2003exploring,
  title={\href{https://doi.org/10.1103/PhysRevLett.90.091301}{Exploring the expansion history of the universe}},
  author={Linder, Eric V},
  journal={Physical review letters},
  volume={90},
  number={9},
  pages={091301},
  year={2003},
  publisher={APS}
}

@article{linden2008test,
  title={\href{https://doi.org/10.1103/PhysRevD.78.023526}{Test of the Chevallier-Polarski-Linder parametrization for rapid dark energy equation of state transitions}},
  author={Linden, Sebastian and Virey, Jean-Marc},
  journal={Physical Review D—Particles, Fields, Gravitation, and Cosmology},
  volume={78},
  number={2},
  pages={023526},
  year={2008},
  publisher={APS}
}

@article{malekjani2025cosmological,
  title={\href{https://doi.org/10.1103/PhysRevD.111.083547}{Cosmological constraints on dark energy parametrizations after DESI 2024: Persistent deviation from standard $\Lambda$ CDM cosmology}},
  author={Malekjani, Mohammad and Davari, Zahra and Pourojaghi, Saeed},
  journal={Physical Review D},
  volume={111},
  number={8},
  pages={083547},
  year={2025},
  publisher={APS}
}

@article{bandyopadhyay2019parameterizing,
  title={\href{https://doi.org/10.1155/2019/5393491}{Parameterizing Dark Energy Models and Study of Finite Time Future Singularities}},
  author={Bandyopadhyay, Tanwi and Debnath, Ujjal},
  journal={Advances in High Energy Physics},
  volume={2019},
  number={1},
  pages={5393491},
  year={2019},
  publisher={Wiley Online Library}
}

@article{pantazis2016comparison,
  title={\href{https://doi.org/10.1103/PhysRevD.93.103503}{Comparison of thawing and freezing dark energy parametrizations}},
  author={Pantazis, George and Nesseris, S and Perivolaropoulos, L},
  journal={Physical Review D},
  volume={93},
  number={10},
  pages={103503},
  year={2016},
  publisher={APS}
}

@article{chaudhary2024early,
  title={\href{https://doi.org/10.1088/1402-4896/ad7178}{Early and late observational tension: dark energy parametrizations in Horava-Lifshitz gravity via baryon acoustic oscillations}},
  author={Chaudhary, Himanshu and Debnath, Ujjal and Rahaman, Farook and Mustafa, G and Atamurotov, Farruh},
  journal={Physica Scripta},
  volume={99},
  number={10},
  pages={105037},
  year={2024},
  publisher={IOP Publishing}
}

@article{rebouccas2024investigating,
  title={\href{https://doi.org/10.48550/arXiv.2408.14628}{Investigating late-time dark energy and massive neutrinos in light of DESI Y1 BAO}},
  author={Rebou{\c{c}}as, Jo{\~a}o and de Souza, Diogo HF and Zhong, Kunhao and Miranda, Vivian and Rosenfeld, Rogerio},
  journal={arXiv preprint arXiv:2408.14628},
  year={2024}
}

@article{jassal2005observational,
  title={\href{https://doi.org/10.1103/PhysRevD.72.103503}{Observational constraints on low redshift evolution of dark energy: How consistent are different observations?}},
  author={Jassal, Harvinder Kaur and Bagla, JS and Padmanabhan, T},
  journal={Physical Review D—Particles, Fields, Gravitation, and Cosmology},
  volume={72},
  number={10},
  pages={103503},
  year={2005},
  publisher={APS}
}

@article{jassal2005wmap,
  title={\href{https://doi.org/10.1111/j.1745-3933.2005.08577.x}{WMAP constraints on low redshift evolution of dark energy}},
  author={Jassal, HK and Bagla, JS and Padmanabhan, T},
  journal={Monthly Notices of the Royal Astronomical Society: Letters},
  volume={356},
  number={1},
  pages={L11--L16},
  year={2005},
  publisher={The Royal Astronomical Society}
}

@article{barboza2008parametric,
  title={\href{https://doi.org/10.1016/j.physletb.2008.08.012}{A parametric model for dark energy}},
  author={Barboza Jr, EM and Alcaniz, JS},
  journal={Physics Letters B},
  volume={666},
  number={5},
  pages={415--419},
  year={2008},
  publisher={Elsevier}
}

@article{chaudhary2024addressing,
  title={\href{https://doi.org/10.1016/j.jheap.2024.08.003}{Addressing the rd tension using late-time observational measurements in a novel deceleration parametrization}},
  author={Chaudhary, Himanshu and Debnath, Ujjal and Maurya, SK and Mustafa, G and Atamurotov, Farruh},
  journal={Journal of High Energy Astrophysics},
  volume={43},
  pages={268--279},
  year={2024},
  publisher={Elsevier}
}

@article{glavan2020einstein,
  title={\href{https://doi.org/10.1103/PhysRevLett.124.081301}{Einstein-Gauss-Bonnet gravity in four-dimensional spacetime}},
  author={Glavan, Dra{\v{z}}en and Lin, Chunshan},
  journal={Physical review letters},
  volume={124},
  number={8},
  pages={081301},
  year={2020},
  publisher={APS}
}

@article{ghosh2020generating,
  title={\href{https://doi.org/10.1088/1361-6382/abc134}{Generating black holes in 4D Einstein--Gauss--Bonnet gravity}},
  author={Ghosh, Sushant G and Kumar, Rahul},
  journal={Classical and Quantum Gravity},
  volume={37},
  number={24},
  pages={245008},
  year={2020},
  publisher={IOP Publishing}
}

@article{zubair2023bouncing,
  title={\href{https://doi.org/10.1140/epjp/s13360-023-03772-1}{Bouncing behaviours in four dimensional Einstein Gauss-Bonnet gravity with cosmography and observational constraints}},
  author={Zubair, M and Farooq, Mushayydha},
  journal={The European Physical Journal Plus},
  volume={138},
  number={2},
  pages={173},
  year={2023},
  publisher={Springer}
}

@article{aoki2020consistent,
  title={\href{https://doi.org/10.1016/j.physletb.2020.135843}{A consistent theory of D→ 4 Einstein-Gauss-Bonnet gravity}},
  author={Aoki, Katsuki and Gorji, Mohammad Ali and Mukohyama, Shinji},
  journal={Physics Letters B},
  volume={810},
  pages={135843},
  year={2020},
  publisher={Elsevier}
}

@article{tsujikawa2022instability,
  title={\href{https://doi.org/10.1016/j.physletb.2022.137329}{Instability of hairy black holes in regularized 4-dimensional Einstein-Gauss-Bonnet gravity}},
  author={Tsujikawa, Shinji},
  journal={Physics Letters B},
  volume={833},
  pages={137329},
  year={2022},
  publisher={Elsevier}
}

@article{banerjee2021quark,
  title={\href{https://doi.org/10.3847/1538-4357/abc87f}{Quark stars in 4D Einstein--Gauss--Bonnet gravity with an interacting quark equation of state}},
  author={Banerjee, Ayan and Tangphati, Takol and Samart, Daris and Channuie, Phongpichit},
  journal={The Astrophysical Journal},
  volume={906},
  number={2},
  pages={114},
  year={2021},
  publisher={IOP Publishing}
}

@article{cardenas2023scalar,
  title={\href{https://doi.org/10.1016/j.nuclphysb.2023.116291}{Scalar-tensor theory with EGB term from Einstein Chern-Simons gravity}},
  author={C{\'a}rdenas, L and Orozco, VC and Salgado, P and Salgado, D and Salgado, R},
  journal={Nuclear Physics B},
  volume={994},
  pages={116291},
  year={2023},
  publisher={Elsevier}
}

@article{PhysRevLett.124.081301,
  title = {\href{https://link.aps.org/doi/10.1103/PhysRevLett.124.081301}{Einstein-Gauss-Bonnet Gravity in Four-Dimensional Spacetime}},
  author = {Glavan, Dra\ifmmode \check{z}\else \v{z}\fi{}en and Lin, Chunshan},
  journal = {Phys. Rev. Lett.},
  volume = {124},
  pages = {081301},
  year = {2020},
  publisher = {American Physical Society},
}

@article{babichev2004black,
  title={\href{https://doi.org/10.1103/PhysRevLett.93.021102}{Black hole mass decreasing due to phantom energy accretion}},
  author={Babichev, Eugeny and Dokuchaev, Vyacheslav and Eroshenko, Yu},
  journal={Physical Review Letters},
  volume={93},
  number={2},
  pages={021102},
  year={2004},
  publisher={APS}
}

@article{babichev2005accretion,
  title={\href{https://doi.org/10.1134/1.1901765}{The accretion of dark energy onto a black hole}},
  author={Babichev, EO and Dokuchaev, VI and Eroshenko, Yu N},
  journal={Journal of Experimental and Theoretical Physics},
  volume={100},
  pages={528--538},
  year={2005},
  publisher={Springer}
}

@article{babichev2013black,
  title={\href{https://doi.org/10.3367/UFNe.0183.201312a.1257}{Black holes in the presence of dark energy}},
  author={Babichev, Eugeniy Olegovich and Dokuchaev, Vyacheslav Ivanovich and Eroshenko, Yu N},
  journal={Physics-Uspekhi},
  volume={56},
  number={12},
  pages={1155},
  year={2013},
  publisher={IOP Publishing}
}

@misc{mukherjee2024accretionphenomenadifferentkinds,
      title={\href{https://doi.org/10.48550/arXiv.2410.20367}{Accretion Phenomena of Different Kinds of Chaplygin Gas Models onto Kehagias-Sfetsos Black Hole in Horava-Lifshitz Gravity Scenario}}, 
      author={Puja Mukherjee and Ujjal Debnath and Anirudh Pradhan},
      year={2024},
      archivePrefix={arXiv},
      primaryClass={gr-qc}, 
}

@article{gonzalez2006some,
  title={\href{https://doi.org/10.1016/j.physletb.2006.02.046}{Some notes on the big trip}},
  author={Gonz{\'a}lez-D{\'\i}az, Pedro F},
  journal={Physics Letters B},
  volume={635},
  number={1},
  pages={1--6},
  year={2006},
  publisher={Elsevier}
}

@article{debnath2020nature,
  title={\href{https://ui.adsabs.harvard.edu/link_gateway/2020GrCo...26..285D/doi:10.1134/S0202289320030056}{Nature of higher-dimensional wormhole mass due to accretion of entropy corrected holographic and new agegraphic dark energies}},
  author={Debnath, Ujjal and Basak, Soumyadipta},
  journal={Gravitation and Cosmology},
  volume={26},
  number={3},
  pages={285--295},
  year={2020},
  publisher={Springer}
}

@article{moresco2012new,
  title={New constraints on cosmological parameters and neutrino properties using the expansion rate of the Universe to z~ 1.75},
  author={Moresco, Michele and Verde, Licia and Pozzetti, Lucia and Jimenez, Raul and Cimatti, Andrea},
  journal={Journal of Cosmology and Astroparticle Physics},
  volume={2012},
  number={07},
  pages={053},
  year={2012},
  publisher={IOP Publishing}
}

@article{moresco2015raising,
  title={Raising the bar: new constraints on the Hubble parameter with cosmic chronometers at z~ 2},
  author={Moresco, Michele},
  journal={Monthly Notices of the Royal Astronomical Society: Letters},
  volume={450},
  number={1},
  pages={L16--L20},
  year={2015},
  publisher={Oxford University Press}
}

@article{moresco20166,
  title={A 6\% measurement of the Hubble parameter at z~ 0.45: direct evidence of the epoch of cosmic re-acceleration},
  author={Moresco, Michele and Pozzetti, Lucia and Cimatti, Andrea and Jimenez, Raul and Maraston, Claudia and Verde, Licia and Thomas, Daniel and Citro, Annalisa and Tojeiro, Rita and Wilkinson, David},
  journal={Journal of Cosmology and Astroparticle Physics},
  volume={2016},
  number={05},
  pages={014},
  year={2016},
  publisher={IOP Publishing}
}

@article{moresco2018setting,
  title={Setting the stage for cosmic chronometers. I. Assessing the impact of young stellar populations on hubble parameter measurements},
  author={Moresco, Michele and Jimenez, Raul and Verde, Licia and Pozzetti, Lucia and Cimatti, Andrea and Citro, Annalisa},
  journal={The Astrophysical Journal},
  volume={868},
  number={2},
  pages={84},
  year={2018},
  publisher={IOP Publishing}
}

@article{moresco2020setting,
  title={Setting the stage for cosmic chronometers. II. Impact of stellar population synthesis models systematics and full covariance matrix},
  author={Moresco, Michele and Jimenez, Raul and Verde, Licia and Cimatti, Andrea and Pozzetti, Lucia},
  journal={The Astrophysical Journal},
  volume={898},
  number={1},
  pages={82},
  year={2020},
  publisher={IOP Publishing}
}

@article{jimenez2002constraining,
  title={Constraining cosmological parameters based on relative galaxy ages},
  author={Jimenez, Raul and Loeb, Abraham},
  journal={The Astrophysical Journal},
  volume={573},
  number={1},
  pages={37},
  year={2002},
  publisher={IOP Publishing}
}

@article{karim2025desi,
  title={Desi dr2 results ii: Measurements of baryon acoustic oscillations and cosmological constraints},
  author={Karim, M Abdul and Aguilar, J and Ahlen, S and Alam, S and Allen, L and Allende Prieto, C and Alves, O and Anand, A and Andrade, U and Armengaud, E and others},
  journal={arXiv e-prints},
  pages={arXiv--2503},
  year={2025}
}

@article{collaboration2020planck,
  title={Planck 2018 results. VI. Cosmological parameters},
  author={Collaboration, Planck and Aghanim, Nabila and Akrami, Y and Ashdown, M and Aumont, J and Baccigalupi, C and Ballardini, M and Banday, AJ and Barreiro, RB and Bartolo, N and others},
  year={2020},
  journal={Astronomy \& Astrophysics},
  publisher={EDP Sciences}
}

@article{pogosian2020recombination,
  title={Recombination-independent determination of the sound horizon and the Hubble constant from BAO},
  author={Pogosian, Levon and Zhao, Gong-Bo and Jedamzik, Karsten},
  journal={The Astrophysical Journal Letters},
  volume={904},
  number={2},
  pages={L17},
  year={2020},
  publisher={IOP Publishing}
}

@article{jedamzik2021reducing,
  title={Why reducing the cosmic sound horizon alone can not fully resolve the Hubble tension},
  author={Jedamzik, Karsten and Pogosian, Levon and Zhao, Gong-Bo},
  journal={Communications Physics},
  volume={4},
  number={1},
  pages={123},
  year={2021},
  publisher={Nature Publishing Group UK London}
}

@article{pogosian2024consistency,
  title={A consistency test of the cosmological model at the epoch of recombination using DESI BAO and Planck measurements},
  author={Pogosian, Levon and Zhao, Gong-Bo and Jedamzik, Karsten},
  journal={arXiv preprint arXiv:2405.20306},
  year={2024}
}

@article{lin2021early,
  title={Early-Universe-Physics Insensitive and Uncalibrated Cosmic Standards: Constraints on $\Omega_{m}$ and Implications for the Hubble Tension},
  author={Lin, Weikang and Chen, Xingang and Mack, Katherine J},
  journal={arXiv preprint arXiv:2102.05701},
  year={2021}
}

@article{vagnozzi2023seven,
  title={Seven hints that early-time new physics alone is not sufficient to solve the Hubble tension},
  author={Vagnozzi, Sunny},
  journal={Universe},
  volume={9},
  number={9},
  pages={393},
  year={2023},
  publisher={MDPI}
}

@article{peebles2003cosmological,
  title={\href{https://doi.org/10.1103/RevModPhys.75.559}{The cosmological constant and dark energy}},
  author={Peebles, P James E and Ratra, Bharat},
  journal={Reviews of modern physics},
  volume={75},
  number={2},
  pages={559},
  year={2003},
  publisher={APS}
}

@article{weinberg1989cosmological,
  title={\href{https://doi.org/10.1103/RevModPhys.61.1}{The cosmological constant problem}},
  author={Weinberg, Steven},
  journal={Reviews of modern physics},
  volume={61},
  number={1},
  pages={1},
  year={1989},
  publisher={APS}
}

@article{padmanabhan2003cosmological,
  title={\href{https://doi.org/10.1016/S0370-1573(03)00120-0}{Cosmological constant—the weight of the vacuum}},
  author={Padmanabhan, Thanu},
  journal={Physics reports},
  volume={380},
  number={5-6},
  pages={235--320},
  year={2003},
  publisher={Elsevier}
}

@article{brout2022pantheon,
  title={The Pantheon+ analysis: cosmological constraints},
  author={Brout, Dillon and Scolnic, Dan and Popovic, Brodie and Riess, Adam G and Carr, Anthony and Zuntz, Joe and Kessler, Rick and Davis, Tamara M and Hinton, Samuel and Jones, David and others},
  journal={The Astrophysical Journal},
  volume={938},
  number={2},
  pages={110},
  year={2022},
  publisher={IOP Publishing}
}

@article{conley2010supernova,
  title={Supernova constraints and systematic uncertainties from the first three years of the supernova legacy survey},
  author={Conley, A and Guy, J and Sullivan, M and Regnault, N and Astier, P and Balland, C and Basa, S and Carlberg, RG and Fouchez, D and Hardin, D and others},
  journal={The Astrophysical Journal Supplement Series},
  volume={192},
  number={1},
  pages={1},
  year={2010},
  publisher={IOP Publishing}
}

@article{trotta2008bayes,
  title={Bayes in the sky: Bayesian inference and model selection in cosmology},
  author={Trotta, Roberto},
  journal={Contemporary Physics},
  volume={49},
  number={2},
  pages={71--104},
  year={2008},
  publisher={Taylor \& Francis}
}

@book{jeffreys1961theory,
  title={The Theory of Probability},
  author={Jeffreys, Harold},
  year={1961},
  publisher={Oxford University Press},
  edition={3rd}
}

@article{copeland2006dynamics,
  title={\href{https://doi.org/10.1142/S021827180600942X}{Dynamics of dark energy}},
  author={Copeland, Edmund J and Sami, Mohammad and Tsujikawa, Shinji},
  journal={International Journal of Modern Physics D},
  volume={15},
  number={11},
  pages={1753--1935},
  year={2006},
  publisher={World Scientific}
}

@article{frieman2008dark,
  title={\href{https://doi.org/10.1146/annurev.astro.46.060407.145243}{Dark energy and the accelerating universe}},
  author={Frieman, Joshua A and Turner, Michael S and Huterer, Dragan},
  journal={Annu. Rev. Astron. Astrophys.},
  volume={46},
  number={1},
  pages={385--432},
  year={2008},
  publisher={Annual Reviews}
}

@article{bagla2003cosmology,
  title={\href{https://doi.org/10.1103/PhysRevD.67.063504}{Cosmology with tachyon field as dark energy}},
  author={Bagla, JS and Jassal, Harvinder Kaur and Padmanabhan, T},
  journal={Physical Review D},
  volume={67},
  number={6},
  pages={063504},
  year={2003},
  publisher={APS}
}

@incollection{tsujikawa2010modified,
  title={\href{https://doi.org/10.1007/978-3-642-10598-2_3}{Modified gravity models of dark energy}},
  author={Tsujikawa, Shinji},
  booktitle={Lectures on cosmology: Accelerated Expansion of the Universe},
  pages={99--145},
  year={2010},
  publisher={Springer}
}

@article{lovelock1971einstein,
  title={\href{https://doi.org/10.1063/1.1665613}{The Einstein tensor and its generalizations}},
  author={Lovelock, David},
  journal={Journal of Mathematical Physics},
  volume={12},
  number={3},
  pages={498--501},
  year={1971},
  publisher={American Institute of Physics}
}

@article{fernandes2020charged,
  title={{Charged black holes in AdS spaces in 4D Einstein Gauss-Bonnet gravity}},
  author={Fernandes, Pedro GS},
  journal={Physics Letters B},
  volume={805},
  pages={135468},
  year={2020},
  publisher={Elsevier}
}

@article{jusufi2020nonlinear,
  title={\href{https://doi.org/10.1016/j.aop.2020.168285}{Nonlinear magnetically charged black holes in 4D Einstein--Gauss--Bonnet gravity}},
  author={Jusufi, Kimet},
  journal={Annals of Physics},
  volume={421},
  pages={168285},
  year={2020},
  publisher={Elsevier}
}

@article{morris1988wormholes,
  title={\href{https://doi.org/10.1119/1.15620}{Wormholes in spacetime and their use for interstellar travel: A tool for teaching general relativity}},
  author={Morris, Michael S and Thorne, Kip S},
  journal={American Journal of Physics},
  volume={56},
  number={5},
  pages={395--412},
  year={1988},
  publisher={American Association of Physics Teachers}
}

@article{morris1988wormhole,
  title={Wormholes, time machines, and the weak energy condition},
  author={Morris, Michael S and Thorne, Kip S and Yurtsever, Ulvi},
  journal={Physical Review Letters},
  volume={61},
  number={13},
  pages={1446},
  year={1988},
  publisher={APS}
}

@article{visser1995lorentzian,
  title={Lorentzian wormholes. from Einstein to Hawking},
  author={Visser, Matt},
  journal={Woodbury},
  year={1995}
}

@article{godani2022stability,
  title={\href{https://doi.org/10.1016/j.dark.2022.100952}{Stability of thin-shell wormhole in 4D Einstein--Gauss--Bonnet gravity}},
  author={Godani, Nisha and Singh, Dharm Veer and Samanta, Gauranga C},
  journal={Physics of the Dark Universe},
  volume={35},
  pages={100952},
  year={2022},
  publisher={Elsevier}
}

@article{hassan2024possibility,
  title={\href{https://doi.org/10.1002/andp.202400114}{Possibility of the traversable wormholes in the galactic halos within 4d Einstein--Gauss--Bonnet gravity}},
  author={Hassan, Zinnat and Sahoo, PK},
  journal={Annalen der Physik},
  volume={536},
  number={8},
  pages={2400114},
  year={2024},
  publisher={Wiley Online Library}
}

@article{john2013accretion,
  title={\href{https://doi.org/10.1103/PhysRevD.88.104005}{Accretion onto a higher-dimensional black hole}},
  author={John, Anslyn J and Ghosh, Sushant G and Maharaj, Sunil D},
  journal={Physical Review D—Particles, Fields, Gravitation, and Cosmology},
  volume={88},
  number={10},
  pages={104005},
  year={2013},
  publisher={APS}
}

@article{debnath2015accretions,
  title={\href{https://doi.org/10.1007/s10509-015-2552-8}{Accretions of dark matter and dark energy onto (n+ 2)-dimensional Schwarzschild black hole and Morris-Thorne wormhole}},
  author={Debnath, Ujjal},
  journal={Astrophysics and Space Science},
  volume={360},
  number={2},
  pages={40},
  year={2015},
  publisher={Springer}
}

@article{debnath2014accretions,
  title={\href{https://doi.org/10.1140/epjc/s10052-014-2869-4}{Accretions of various types of dark energies onto Morris-Thorne wormhole}},
  author={Debnath, Ujjal},
  journal={The European Physical Journal C},
  volume={74},
  pages={1--8},
  year={2014},
  publisher={Springer}
}

@article{ade2014planck,
  title={\href{https://doi.org/10.1051/0004-6361/201321529}{Planck 2013 results. I. Overview of products and scientific results}},
  author={Ade, Peter AR and Aghanim, Nabila and Alves, MIR and Armitage-Caplan, Charmaine and Arnaud, M and Ashdown, M and Atrio-Barandela, F and Aumont, J and Aussel, H and Baccigalupi, C and others},
  journal={Astronomy \& Astrophysics},
  volume={571},
  pages={A1},
  year={2014},
  publisher={EDP sciences}
}

@article{adam2016planck,
  title={\href{https://doi.org/10.1051/0004-6361/201527101}{Planck 2015 results-I. Overview of products and scientific results}},
  author={Adam, R{\'e}mi and Ade, Peter AR and Aghanim, N and Akrami, Y and Alves, MIR and Arg{\"u}eso, F and Arnaud, M and Arroja, F and Ashdown, Mark and Aumont, J and others},
  journal={Astronomy \& Astrophysics},
  volume={594},
  pages={A1},
  year={2016},
  publisher={EDP sciences}
}

@article{guy2010supernova,
  title={\href{https://doi.org/10.1051/0004-6361/201014468}{The Supernova Legacy Survey 3-year sample: Type Ia supernovae photometric distances and cosmological constraints}},
  author={Guy, Julien and Sullivan, Mark and Conley, A and Regnault, N and Astier, P and Balland, C and Basa, S and Carlberg, RG and Fouchez, D and Hardin, D and others},
  journal={Astronomy \& Astrophysics},
  volume={523},
  pages={A7},
  year={2010},
  publisher={EDP Sciences}
}

@article{li2013planck,
  title={\href{https://doi.org/10.1088/1475-7516/2013/09/021}{Planck constraints on holographic dark energy}},
  author={Li, Miao and Li, Xiao-Dong and Ma, Yin-Zhe and Zhang, Xin and Zhang, Zhenhui},
  journal={Journal of Cosmology and Astroparticle Physics},
  volume={2013},
  number={09},
  pages={021},
  year={2013},
  publisher={IOP Publishing}
}

@article{adame2025desi,
  title={\href{https://doi.org/10.1088/1475-7516/2025/02/021}{DESI 2024 VI: Cosmological constraints from the measurements of baryon acoustic oscillations}},
  author={Adame, AG and Aguilar, J and Ahlen, S and Alam, S and Alexander, DM and Alvarez, M and Alves, O and Anand, A and Andrade, U and Armengaud, E and others},
  journal={Journal of Cosmology and Astroparticle Physics},
  volume={2025},
  number={02},
  pages={021},
  year={2025},
  publisher={IOP Publishing}
}

@article{adame2025desis,
  title={\href{https://doi.org/10.1088/1475-7516/2025/04/012}{DESI 2024 III: Baryon acoustic oscillations from galaxies and quasars}},
  author={Adame, AG and Aguilar, J and Ahlen, S and Alam, S and Alexander, DM and Alvarez, M and Alves, O and Anand, A and Andrade, U and Armengaud, E and others},
  journal={Journal of Cosmology and Astroparticle Physics},
  volume={2025},
  number={04},
  pages={012},
  year={2025},
  publisher={IOP Publishing}
}

@article{adame2025desii,
  title={\href{https://doi.org/10.1088/1475-7516/2025/01/124}{DESI 2024 IV: Baryon acoustic oscillations from the Lyman alpha forest}},
  author={Adame, AG and Aguilar, J and Ahlen, S and Alam, S and Alexander, DM and Alvarez, M and Alves, O and Anand, A and Andrade, U and Armengaud, E and others},
  journal={Journal of Cosmology and Astroparticle Physics},
  volume={2025},
  number={01},
  pages={124},
  year={2025},
  publisher={IOP Publishing}
}

@article{paul2013observational,
  title={\href{https://doi.org/10.1007/s12043-013-0593-5}{Observational constraints on modified Chaplygin gas in Horava--Lifshitz gravity with dark radiation}},
  author={Paul, BC and Thakur, P and Verma, MM},
  journal={Pramana},
  volume={81},
  pages={691--718},
  year={2013},
  publisher={Springer}
}

@article{paul2012modified,
  title={\href{https://doi.org/10.1103/PhysRevD.85.024039}{Modified Chaplygin gas in Horava-Lifshitz gravity and constraints on its B parameter}},
  author={Paul, BC and Thakur, P and Saha, A},
  journal={Physical Review D—Particles, Fields, Gravitation, and Cosmology},
  volume={85},
  number={2},
  pages={024039},
  year={2012},
  publisher={APS}
}

@article{basak2025accretion,
  title={\href{https://doi.org/10.1140/epjc/s10052-025-14398-1}{Accretion of dark energy onto black hole in Bumblebee field}},
  author={Basak, Anuka and Debnath, Ujjal},
  journal={The European Physical Journal C},
  volume={85},
  number={6},
  pages={1--17},
  year={2025},
  publisher={Springer}
}

@article{mukherjee2024constraining,
  title={\href{https://ui.adsabs.harvard.edu/link_gateway/2024EPJC...84..930M/doi:10.1140/epjc/s10052-024-13196-5}{Constraining the parameters of generalized and viscous modified Chaplygin gas and black hole accretion in Einstein-Aether gravity}},
  author={Mukherjee, Puja and Debnath, Ujjal and Chaudhary, Himanshu and Mustafa, G},
  journal={The European Physical Journal C},
  volume={84},
  number={9},
  pages={930},
  year={2024},
  publisher={Springer}
}

@article{mukherjee2025parameter,
  title={\href{https://doi.org/10.1088/1475-7516/2025/05/085}{How parameter constraining can influence the mass accretion process of a black hole in the generalized Rastall gravity theory?}},
  author={Mukherjee, Puja and Debnath, Ujjal and Chaudhary, Himanshu and Mustafa, G},
  journal={Journal of Cosmology and Astroparticle Physics},
  volume={2025},
  number={05},
  pages={085},
  year={2025},
  publisher={IOP Publishing}
}

@article{mukherjee2023accretion,
  title={\href{https://doi.org/10.1142/S0219887823502183}{Accretion of modified Chaplygin-Jacobi gas and modified Chaplygin-Abel gas onto Schwarzschild black hole}},
  author={Mukherjee, Puja and Debnath, Ujjal and Pradhan, Anirudh},
  journal={International Journal of Geometric Methods in Modern Physics},
  volume={20},
  number={12},
  pages={2350218},
  year={2023},
  publisher={World Scientific}
}

@article{huang2009fitting,
  title={\href{https://doi.org/10.1103/PhysRevD.80.083515}{Fitting the constitution type Ia supernova data with the redshift-binned parametrization method}},
  author={Huang, Qing-Guo and Li, Miao and Li, Xiao-Dong and Wang, Shuang},
  journal={Physical Review D—Particles, Fields, Gravitation, and Cosmology},
  volume={80},
  number={8},
  pages={083515},
  year={2009},
  publisher={APS}
}

@article{feng2012new,
  title={\href{https://doi.org/10.1088/1475-7516/2012/09/023}{A new class of parametrization for dark energy without divergence}},
  author={Feng, Chao-Jun and Shen, Xian-Yong and Li, Ping and Li, Xin-Zhou},
  journal={Journal of Cosmology and Astroparticle Physics},
  volume={2012},
  number={09},
  pages={023},
  year={2012},
  publisher={IOP Publishing}
}

@article{liu2008revisiting,
  title={\href{https://doi.org/10.1111/j.1365-2966.2008.13380.x}{Revisiting the parametrization of equation of state of dark energy via SNIa data}},
  author={Liu, Dao-Jun and Li, Xin-Zhou and Hao, Jiangang and Jin, Xing-Hua},
  journal={Monthly Notices of the Royal Astronomical Society},
  volume={388},
  number={1},
  pages={275--281},
  year={2008},
  publisher={The Royal Astronomical Society}
}

\end{document}